\documentclass[useAMS,usenatbib,usegraphicx]{mn2e}

\def\gtrsim{\mathrel{\hbox{\rlap{\hbox{\lower3pt\hbox{$\sim$}}}\hbox{\raise2pt\hbox{$>$}}}}}

\title[Unveiling new members in nearby young moving groups]{Unveiling new
members in five nearby young moving groups}
\author[A. Mo\'or et al.]{A. Mo\'or$^{1}$\thanks{E-mail:moor@konkoly.hu}, 
Gy.~M. Szab\'o$^{1,2,3}$, L.~L. Kiss$^{1,2,4}$, Cs. Kiss$^{1}$, P. \'Abrah\'am$^{1}$, J. Szul\'agyi$^{1}$\\
\newauthor \'A. K\'osp\'al$^{5}$, T. Szalai$^{3}$\\
$^{1}$Konkoly Observatory, Research Centre for Astronomy and Earth Sciences, \\ 
Hungarian Academy of Sciences, PO Box 67, 
H-1525 Budapest, Hungary\\
$^{2}$ELTE Gothard-Lend\"ulet Research Group, 9700 Szombathely, Hungary\\
$^{3}$Dept. of Experimental Physics and Astronomical Observatory, 6720 Szeged D\'om t\'er 9., Hungary\\
$^{4}$Sydney Institute for Astronomy, School of Physics, University of Sydney, NSW 2006, Australia\\
$^{5}$Research and Scientific Support Department, European Space Agency\\ 
(ESA-ESTEC, SRE-SA),\\ 
 P.O. Box 299, 2200 AG, Noordwijk, The Netherlands ; ESA fellow.\\
}

\begin{document}

\date{Accepted ... Received ...; in original form ...}

\pagerange{\pageref{firstpage}--\pageref{lastpage}} \pubyear{2013}

\maketitle

\label{firstpage}

\begin{abstract}
In the last decade many kinematic groups of young stars ($<$~100~Myr) were discovered in the solar
neighbourhood. Since the most interesting period of planet formation overlaps with the 
age of these groups, their well dated members are attractive targets for exoplanet searches by direct imaging. 
{We combined astrometric, photometric and X-ray data, and applied strict selection criteria to explore the stellar 
content of five nearby moving groups. We identified more than 100 potential new candidate members in the 
$\beta$\,Pic moving group, and in the Tucana-Horologium, Columba, Carina, and Argus associations.}
%With the aim to explore the stellar content of nearby young moving groups further,
%we identified more than 100 potential new candidate members in five kinematic groups 
%($\beta$ Pic moving group, Tucana-Horologium, Columba, Carina, and Argus associations), 
%by combining astrometric, photometric, X-ray data and applying strict selection criteria. 
In order to further assess and confirm their membership status, we analysed 
radial velocity data and lithium equivalent widths extracted from 
high-resolution spectra of 54 candidate stars. 
We identified 
35 new probable/possible young moving group members: 4 in the $\beta$\,Pic moving group, 
11 in the Columba association, 16 in the Carina association, and 4 in the Argus association. 
We found serendipitiously a new AB~Dor moving group member as well.
For four Columba systems {\sl Hipparcos} based parallaxes 
have already been available
 and as they are consistent with the predicted kinematic parallaxes, they can be considered 
as secure new members.
\end{abstract}


\section{Introduction}

Nearby young moving groups are gravitationally unbound, loose associations of stars 
that have common origin and {move} through space together. These groups 
occupy extended regions ($\sim$100~pc) in our Solar System's neighbourhood and due to their 
proximity the group members appear to us widely spread across the sky.
This makes the  
integrity of the systems hardly recognizable.
Members of a specific group can be linked together  
through their common space velocities and common ages. Such a study requires 
the combination of astrometric data 
with radial velocity information and application of relevant age diagnostic methods.
This is why the intensive investigation 
of nearby young moving groups started only
in the late nineties of the last century, after the {\sl Hipparcos} and {\sl TYCHO} all-sky 
astrometric catalogues and the {\sl ROSAT} all-sky point source catalogues (X-ray activity can be used 
as a youth indicator) became available. 
Thanks to subsequent studies, up to now nine young ($<$100~Myr) kinematic groups have been identified 
in the vicinity of the Sun \citep[for a review, see][]{zucksong2004,torres2008}.

According to the review by \citet{torres2008}, more than 
300 members of these nine groups are known. 
The census of these groups is far from complete, because the necessary kinematic information, 
particularly radial velocities and trigonometric distances, 
{are} missing for most of the nearby stars.
This is especially true for low-mass populations of moving groups, whose census may be 
highly incomplete \citep[e.g.][]{shkolnik2009}.  
Thanks to studies focused on searching for such young low mass stars, {a} large number of new 
M-type group members/candidates were identified in the last few 
years \citep{malo2013,schlieder2010,schlieder2012a,schlieder2012b,shkolnik2011,shkolnik2012}.
The census of the high mass regime could, {however,} be incomplete as well, as demonstrated by recent discoveries 
of several new early-type members of these groups \citep{zuckerman2011,zuckerman2012}. 

A significant fraction of nearby stars 
younger than 100~Myr belongs to the known kinematic groups.
The ages of these nearby young moving groups range between $\sim$6 and 100 million year, 
thus -- according to the current paradigm -- overlap with the most interesting 
period of planetary system formation.  Gas giant planets must form before the gas content 
of the primordial protoplanetary disc is largely dissipated 
\citep[i.e. within 10~Myr,][]{pascucci2006}, 
while terrestrial planets may reach their
final mass {on time-scales of} ten millions of years \citep{chambers2001,kb2006}.
Although the formation
of planetesimals and planetesimal belts 
in the outer region of planetary systems may {even last} for hundred million years, the most
active period is restricted to the first few tens of million years \citep{kb2008}. 
The evolution of planetesimal belts can be traced via the investigation of debris 
disks that are linked to them \citep{wyatt2008}.
Thus, members of nearby young moving groups are
suitable targets to study the evolution of planetary systems, particularly 
because, unlike field stars, their ages are well determined. 
These young kinematic groups also offer insights into the star formation process in 
low-density environments \citep{fernandez2008}.

Young stars 
are excellent targets to detect sub-stellar objects via direct imaging,
since giant gas planets are 
thought to have significant excess luminosity due to the rapid contraction 
in the early phase of their evolution \citep{kasper2007}. 
Indeed,
HR~8799 -- a star with a multiple planetary system
 discovered via direct imaging \citep{marois2008} -- may belong to 
 the 30~Myr old Columba association. 
 $\kappa$~And, hosting star of a super-Jupiter successfully imaged with 
Subaru/HiCIAO,   
can also be assigned to the Columba association \citep{carson2012}.
 $\beta$~Pic, the member and namesake of 
the $\beta$~Pic moving group, also harbours an imaged planet \citep{lagrange2010}.

In our study, we searched for new members in 
five young assemblages, the $\beta$\,Pic moving group (BPMG), the Tucana-Horologium  (THA),  
the Columba (COL), the Carina (CAR), and the Argus (ARG) associations. 
Each selected kinematic group is younger than 50\,Myr and most of their known members
are located within 120~pc to the Sun. In Appendix~\ref{appendix}, we briefly summarise the main properties of 
the selected young associations.
We selected potential candidate members from a catalogue compiled by cross-correlating 
the entries of the Naval Observatory 
Merged Astrometric Dataset \citep[NOMAD,][]{zacharias2004} with the 
entries of the 
{\sl ROSAT} All-Sky Bright Source Catalogue and the
{\sl ROSAT} All-Sky Survey Faint Source Catalog \citep{voges1999,voges2000} (Sect.~\ref{selection}).
Due to the selection method {applied} our survey was most efficient in 
finding F5-M type 
group members.
 In order to confirm the membership of our potential candidates, we performed 
follow-up high-resolution spectroscopy (Sect.~\ref{obsanddatared}).
The analysis of these spectra and the final assignments of the candidate stars 
are reviewed in Sect.~\ref{results}, our results are summarized in 
Sect.~\ref{summary}. 

%%%%%%%%%%%%%%%%%%%%%%%%%%%%%%%%%%%%%%%%%%%%%%%%%%%%%%%%%%%%%%%%%%%%%%%%%%%%%%
%%%%%%%%%%%%%  CANDIDATE SELECTION

\section{Candidate selection} 

\subsection{Selection method} \label{selection}

Young stars with spectral type F and later are 
known to exhibit enhanced coronal activity with 
strong X-ray emission, making the latter property a good indicator 
of youth \citep{zucksong2004}. Exploiting this characteristic as a first step of our 
survey, we cross-correlated the entries of the 
NOMAD catalogue
with the 
entries of {\sl ROSAT} catalogues
 compiling an initial list of potential young stars from the whole sky. 
For real coincidences we demanded that the NOMAD counterpart(s) 
must be located within a distance of less than 2 times the {\sl ROSAT} positional error
from the X-ray source's position. We included only those stars 
whose proper motion measurement fulfilled the following criteria:
1) $\rm \mu = \sqrt{ {\mu_{\alpha}}^2 {\cos{\delta}}^2 + {\mu_{\delta}}^2 } 
> 20$\,mas yr$^{-1}$; 2) $\rm {\mu}/\sigma_{\mu}>$5.
Because of the selection method, the compiled sample contains primarily 
 stars with spectral type F5 or later.
 In order to select the most probable
candidate members of the chosen five young moving groups, this starting sample 
of likely young stars was further analysed. 
 
Stars {that belong} to a specific kinematic 
group share common spatial motion through the Milky Way. 
Due to projection effects, the common space motion manifests itself as a converging 
pattern, and the motion of the members defines a common point of convergence in the 
sky. Thus, based on positional and proper motion data, one can 
isolate possible members of a group. We used the proper motion selection
algorithm proposed by \citet{lepine2009} to identify objects 
with proper motions consistent with membership in any of our five selected groups.
Positional and proper motion data of candidate sources 
were based on the NOMAD catalogue, the characteristic galactic space 
motion of selected young moving groups 
was taken from \citet{torres2008}. 
For each of our candidates, we calculated the projected motion of 
all five groups in the plane of the sky local to the
star and then computed the 
$\Phi$ angle that these vectors 
subtend with the candidate's proper motion vector 
\citep[Eq.~1--3 in][]{lepine2009}. 
The smaller the value of $\Phi$ the better the consistency 
between the motion of the candidate star and the characteristic 
motion of a kinematic group. In order to determine an upper threshold for 
 $\Phi$, we repeated 
these calculations for the known members of the selected five 
groups \citep[the memberlist was taken from][]{torres2008}. We found that 
for more than 90\% of known members
the value of $\Phi$ is less than $12\degr$. Adopting  
this value
as an upper threshold we omitted all those stars from the starting sample 
whose proper motion was not consistent with the mean projected proper motion 
of any of the selected moving groups. 
Assuming that a certain star belongs to a specific kinematic group, 
we estimated a kinematic 
parallax for all of 
the remaining candidates and 
predicted their radial velocity as well \citep[Eq.~6 and 8 in][]{lepine2009}.

To refine the list of candidates further, we derived K-band absolute magnitudes
from the kinematic distances ($M_{Ks,kin}$) or from the trigonometric distances ($M_{Ks,tri}$) 
{when} {\sl Hipparcos} data were available. These were compared with the absolute photometric K 
magnitudes ($M_{Ks,pho}$) calculated from observational isochrones.
The observational isochrones were defined 
using the colour-magnitude diagram ($M_{Ks}$ vs. $V-K_s$) of 
the specific groups where the locus of the known members was
fitted with a linear relationship (see e.g. Fig.~\ref{cmd}). The members of the co-eval THA, COL, and CAR groups 
were fitted synchronously. For true moving group members, $M_{Ks,kin}$ (or $M_{Ks,tri}$)
and $M_{Ks,pho}$ must be consistent. Based on the known members we found 
that $|\Delta M_{Ks}| = |M_{Ks,kin/tri} - M_{Ks,pho}| \leq 1$~mag, therefore, 
candidate objects with 
$|\Delta M_{Ks}| > 1$~mag were discarded from the sample. 

Using the kinematic distances, we computed the candidates' physical space 
coordinates centred on the Sun (X,Y,Z) and compared them to the region defined 
by the known members of the assigned group.
In most cases the boundaries of a specific group's region (ranges in 
X, Y, Z coordinates) were taken from \citet[][see their Table~2]{torres2008}.
In \citet{moor2011b} we reported the discovery of a new probable Columba 
member, {HD~3670}, which is located slightly out of the 
region defined in the abovementioned work. 
Taking into account the position of this new member 
for Columba, we adopted a region that is somewhat more extended 
in the X direction (X$=-106 - +20$~pc) than that defined in 
\citet[][X$=-106 - +9$~pc]{torres2008}.  
In the following, we ignore candidates located {outside} 
the defined regions.

\begin{figure} 
\includegraphics[width=85mm]{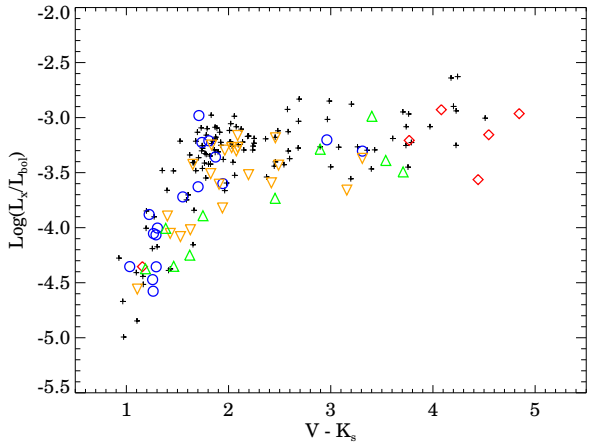}
\caption{Fractional X-ray luminosities as a function of V$-$K$_{\rm s}$ for 
the known and candidate members of the five kinematic groups. 
Plus signs represent the previously known members of the groups.
Candidate members of BPMG, COL, CAR and ARG are taken from 
Table~\ref{stellarprop} and marked by 
diamonds (red in the electronic version), circles (blue in the electronic 
version),  
upside down triangles (orange in the electronic version) and 
triangles (green in the electronic version), respectively.
}
\label{rxs}
\end{figure}

Our sample consists of stars that have X-ray counterparts. However, even 
quite old stars can exhibit X-ray emission and could be detected by {\sl ROSAT}.
Rotation and activity of sun-like main sequence stars  
diminish during their life.
Potential young star can be selected based on their enhanced coronal activity 
that is observable in their strong X-ray emission and high fractional X-ray 
luminosity (${L_{\rm x}}/{L_{\rm bol}}$). 
We plotted fractional X-ray luminosities of candidate stars
as a function of their $V-K_s$ colour indices
and compared them with the distribution of known members in the 
same plot. Since the X-ray activity of known members of the selected groups
as a function of $V-K_s$ are very similar to each other, in this case we used 
them as an ensemble of young stars and did not separate them based on their 
group membership. 
After this comparison we kept only those candidates 
whose X-ray properties were consistent with that of known members 
(see e.g. Fig.~\ref{rxs}).

As a final step of our selection process, we searched the literature 
for additional data on our candidate stars and rejected
those targets that 1) have {already been} identified as a member 
of a young moving group; 2) are included in the {\sl Hipparcos} 
catalogue and their measured trigonometric parallaxes are 
not consistent with the kinematic ones; 3) have radial velocity data 
inconsistent with the predicted ones; 4) reside in a binary 
system, where the companion has a measured trigonometric parallax and/or 
radial velocity which is/are inconsistent with the 
predicted value(s). As a result of the applied selection algorithm,
we finally identified 129 stars that could potentially be
assigned to at least one of the selected five young moving groups.
There are several stars which could be assigned 
to more than one group. 

Members of a young moving group are expected to share {a} similar space 
motion.
Consistency between the proper motion of 
a candidate and the projected motion of a moving group is a 
necessary but not sufficient condition of the membership, {since e.g.} 
stars can move in the same
direction but with a significantly different speed.
The computation of the heliocentric space motion (U,V,W) {of} a star 
requires the knowledge of {both} its radial velocity and distance. 
X-ray activity of stars measured by their fractional X-ray luminosity 
is a useful youth indicator, however, it is not applicable as an 
accurate age diagnostic for 10--100\,Myr old stars since
in this range this property shows {a} weak dependency on age.  
By measuring the radial velocity of the candidates and by applying additional  
age diagnostic methods (e.g. lithium content),
we can collect further evidences for their membership.

We obtained high-resolution spectra for fifty candidate stars 
from the compiled list. For four additional objects we 
used data from the SACY survey \citep{torres2006} 
to complement our observed sample. 
Hereafter, throughout this paper, we focus on the analysis
of these 54 stars.

\subsection{Basic properties of the selected stars}

Basic properties of the 54 stars as well as their suspected host kinematic groups 
are summarized in Table~\ref{stellarprop}. 
There are several stars for which more than one kinematic groups are listed. 
We note that for these 54 targets, 
we also investigated the possibility whether they could be the member 
of three additional groups: {the} TW Hya association, {the} AB Doradus moving group, {and the}
$\eta$ Cha group. 
In Table~\ref{stellarprop}, for a certain candidate star, we list the relevant groups 
in ascending {order of $\Phi$ values} 
(smaller $\Phi$ 
indicates better consistency 
{of} the motion of the candidate star {with} the characteristic 
motion of a kinematic group).

In order to estimate the effective temperature of our targets, 
we modelled the stellar photosphere by fitting the optical and 
infrared photometric data with a {\sc NEXTGEN} atmosphere model 
\citep{hauschildt}.
Optical photometry, when available, were taken from the {\sl TYCHO2} catalogue,  
 otherwise we used the $V$ magnitude quoted in the NOMAD catalogue.
Additional near-infrared data in $J, H, K_s$ bands 
were taken from the Two Micron All-Sky Survey \citep[2MASS;][]{skrutskie}.
This data set was further supplemented 
by Wide-field Infrared Survey Explorer ({\sl WISE}) $W1$ band (centred {at} 
3.4{\micron}) 
photometry from the {\sl WISE} All-Sky Database \citep{wright}.
The surface gravity was fixed to $\log{g}=$4.5~cms$^{-2}$ 
and 
we adopted solar metallicity for all 
of our targets. 
Since all of our stars are predicted to be located within 125\,pc 
\citep[i.e. most of them can be found inside the Local Bubble, where the mean extinction is low,][]{lallement2003}
the visual extinction was neglected in the fitting process. 
The $T_{\rm eff}$ values obtained are listed in Table~\ref{stellarprop}.

%%%%%%%%%%%%%%%%%%%%%%%%%%%%%%%%%%%%%%%%%%%%%%%%%%%%%%%%%%%%%%%%%%%%%%%%%%%%%%
%%%%%%%%%%%%%  TABLE1

%%%%%%%%%%%%%%%%%%%%%%%%%%%%%%%%%%%%%%%%%%%%%%%%%%%%%%%%%%%%%%%%%%%%%%%%%%%%%%
%\input{tables_v2/table1.tex}
%\input{table1.tex}
\begin{table*}                                                                  
\setlength{\tabcolsep}{1.6mm}                                                   
\begin{center}                                                                  
\scriptsize                                                                     
\caption{Basic properties of candidate members.         
(1) NOMAD-identifier of the star. (2) Other, SIMBAD compatible, name of the     
star. (3) Right ascension (J2000). (4) Declination                              
(J2000). (5) {Spectral type, taken from SIMBAD if available, 
otherwise (marked by asterisks) estimated from the derived effective temperature.
For the latter calculation we used the conversion tables from \citet{pecaut2013}:
for stars that we 
 identified as a probable member of young moving groups (see Sect.~\ref{finalassignment})
we used their table~4 (with T$_{\rm eff}$-spectral type tabulation for 5--30~Myr pre-main-sequence stars), 
in other cases the T$_{\rm eff}$ and spectral type data were taken from their table~6 (tabulation for dwarfs).}
(6) 2MASS K$_s$ magnitude.       
(7) V$-$K$_s$ colour index. (8) Derived effective temperature.                  
(9) Fractional X-ray luminosity based on ROSAT data.                            
0466-0044732, 0466-0044706 and 0601-0968580 are members of binary systems where 
both components are within the positional uncertainty circle of {a} nearby      
ROSAT source.                                                                   
To account for binarity, in these cases the X-ray flux was divided          
by two.                                                                         
%when comparing to the X-ray properties of known moving group members.          
(10) Identifiers of candidate moving groups.                                    
There are several stars that could be assigned to more than one group.          
 In these cases the groups                                                      
 are listed in increasing {order of $\Phi$ values}.                               
\label{stellarprop}
}                                                           
\begin{tabular}{cccccccccc}                                                     
\hline\hline                                                                    
    NOMAD ID &                Other name &    RA (2000) &   DEC (2000) &      SpT &    K$_s$ & V--K$_s$ &   T$_{eff}$ &    $\log{\frac{L_{\rm x}}{L_{\rm bol}}}$ & Candidate groups \\
             &                           &              &              &          &    [mag] &    [mag] &         [K] &                                          &                  \\
         (1) &                       (2) &          (3) &          (4) &      (5) &      (6) &      (7) &         (8) &                                      (9) &             (10) \\
\hline
0367-0004557 &                   HD 3888 &   00:40:51.6 &  -53:12:35.7 &      F7V &     6.11 &     1.26 &        6250 &    -4.47 &          COL/THA/ABD \\
0361-0008966 &                   HD 8077 &   01:19:05.6 &  -53:51:01.9 &      F6V &     7.61 &     1.27 &        6200 &    -4.05 &                  COL \\
0540-0027893 &                CD-36 1289 &   03:25:51.9 &  -35:56:25.5 &      F8V &     8.14 &     1.74 &        5400 &    -3.23 &                  COL \\
0453-0032480 &                CD-45 1384 &   04:03:53.4 &  -44:39:31.6 &   G5$^*$ &     8.51 &     1.71 &        5450 &    -2.98 &                  COL \\
0454-0034540 &                CD-44 1533 &   04:22:45.7 &  -44:32:51.9 &      K0V &     8.58 &     1.87 &        5400 &    -3.36 &                  COL \\
0466-0044706 &                 HIP 25434 &   05:26:23.0 &  -43:22:36.3 &       G0 &     7.76 &     1.29 &        6350 &    -4.07 &                  COL \\
0466-0044732 &                 HIP 25436 &   05:26:24.0 &  -43:22:32.7 &   F4IV/V &     7.21 &     1.03 &        6700 &    -4.35 &                  COL \\
0211-0114754 &                 HD 269620 &   05:29:27.1 &  -68:52:04.9 &      G6V &     8.06 &     1.55 &        5700 &    -3.72 &                  COL \\
0298-0039269 &                  HD 37402 &   05:34:26.1 &  -60:06:15.1 &      F6V &     7.09 &     1.29 &        6200 &    -4.36 &                  COL \\
0211-0129478 &                 HD 269921 &   05:38:34.5 &  -68:53:06.6 &      G7V &     8.39 &     1.81 &        5400 &    -3.21 &                  COL \\
0404-0048765 &                  HD 38470 &   05:43:24.6 &  -49:31:06.4 &      F4V &     8.03 &     1.19 &        6400 &    -4.37 &                  ARG \\
0351-0047432 &                  HD 39130 &   05:47:18.0 &  -54:50:28.2 &      F6V &     7.43 &     1.26 &        6250 &    -4.58 &                  COL \\
0287-0048860 &                  HD 44345 &   06:17:20.9 &  -61:14:46.7 &      F5V &     7.39 &     1.11 &        6450 &    -4.56 &                  CAR \\
0339-0066782 &           TYC 8554-1187-1 &   07:11:57.3 &  -56:04:21.5 &   F7$^*$ &     8.97 &     1.22 &        6250 &    -3.88 &                  COL \\
0315-0057897 &                CD-58 1720 &   07:20:27.9 &  -58:29:26.6 &   G3$^*$ &     8.44 &     1.63 &        5700 &    -4.02 &                  CAR \\
0352-0070710 &               CPD-54 1295 &   07:31:19.9 &  -54:47:02.8 &       G5 &     8.75 &     1.75 &        5500 &    -3.89 &                  ARG \\
0368-0077090 &                         - &   07:39:12.2 &  -53:07:29.8 &   K6$^*$ &     8.75 &     3.31 &        4200 &    -3.30 &                  COL \\
0263-0084896 &                 CD-63 336 &   07:42:38.9 &  -63:36:14.0 &   G2$^*$ &     8.39 &     1.53 &        5850 &    -4.08 &              CAR/COL \\
0664-0176864 &                CD-23 6751 &   08:04:24.8 &  -23:35:00.2 &       K0 &     8.16 &     1.70 &        5500 &    -3.63 &                  COL \\
0331-0114466 &                         - &   08:09:42.7 &  -56:52:20.1 &   K0$^*$ &     8.83 &     2.08 &        5050 &    -3.30 &                  CAR \\
0406-0105889 &             TYC 8157-91-1 &   08:15:21.6 &  -49:18:30.3 &   G7$^*$ &     9.11 &     1.94 &        5300 &    -3.60 &                  COL \\
0248-0110675 &           TYC 8933-1204-1 &   08:20:56.0 &  -65:10:12.4 &   K2$^*$ &     8.93 &     2.49 &        4750 &    -3.43 &              CAR/COL \\
0248-0115043 &                  HD 73129 &   08:32:46.2 &  -65:08:27.2 &      G2V &     7.85 &     1.46 &        5850 &    -4.35 &              ARG/COL \\
0354-0106699 &            TYC 8573-579-1 &   08:41:01.3 &  -54:31:02.3 &   G1$^*$ &     9.01 &     1.39 &        5950 &    -4.01 &                  ARG \\
0341-0108561 &                         - &   08:44:25.2 &  -55:48:47.8 &   K3$^*$ &     9.56 &     2.90 &        4450 &    -3.29 &                  ARG \\
0489-0152365 &                CD-40 4610 &   08:46:46.9 &  -41:05:46.7 &   F8$^*$ &     8.69 &     1.30 &        6100 &    -4.00 &                  COL \\
0269-0122275 &                         - &   08:52:43.6 &  -63:04:18.0 &   G3$^*$ &     8.68 &     1.65 &        5700 &    -3.42 &                  CAR \\
0291-0099052 &           TYC 8927-2869-1 &   08:59:04.5 &  -60:51:46.6 &   G9$^*$ &     9.19 &     2.19 &        5100 &    -3.52 &                  CAR \\
0372-0169783 &                         - &   09:00:59.3 &  -52:44:45.4 &   K4$^*$ &     9.42 &     2.96 &        4400 &    -3.20 &              COL/CAR \\
0313-0102916 &           TYC 8594-1049-1 &   09:01:45.6 &  -58:41:06.2 &   K0$^*$ &     9.10 &     2.09 &        5000 &    -3.27 &              COL/CAR \\
0397-0143387 &           TYC 8174-1586-1 &   09:11:15.8 &  -50:14:14.9 &      K5e &     9.50 &     2.46 &        4650 &    -3.18 &              CAR/COL \\
0268-0125760 &            TYC 8944-874-1 &   09:11:25.3 &  -63:11:28.2 &   K2$^*$ &     8.55 &     2.03 &        5050 &    -3.27 &          COL/ARG/CAR \\
0348-0138354 &           TYC 8587-1015-1 &   09:15:14.9 &  -55:11:51.6 &   K0$^*$ &     9.13 &     1.96 &        5300 &    -3.30 &              CAR/COL \\
0308-0116109 &           TYC 8595-1740-1 &   09:15:57.6 &  -59:11:07.1 &   G5$^*$ &     9.22 &     1.83 &        5450 &    -3.25 &              CAR/COL \\
0337-0139009 &            TYC 8591-145-1 &   09:17:00.5 &  -56:16:48.1 &   G9$^*$ &     9.08 &     1.94 &        5400 &    -3.82 &              COL/CAR \\
0415-0155444 &                  HD 80595 &   09:19:04.2 &  -48:24:52.2 &      G0V &     8.17 &     1.40 &        6000 &    -3.89 &              CAR/COL \\
0193-0211246 &            TYC 9200-446-1 &   09:30:31.5 &  -70:41:48.0 &   G5$^*$ &     8.56 &     1.83 &        5500 &    -3.51 &              CAR/ARG \\
0367-0179130 &                         - &   09:31:25.4 &  -53:14:36.6 &   K5$^*$ &     9.50 &     3.31 &        4200 &    -3.37 &              CAR/COL \\
0393-0168117 &                         - &   09:36:58.6 &  -50:41:56.1 &   K2$^*$ &     9.46 &     2.09 &        5000 &    -3.16 &              COL/CAR \\
0327-0202270 &                 HD 302321 &   09:41:25.6 &  -57:15:22.0 &       F8 &     8.42 &     1.43 &        5950 &    -4.05 &              CAR/COL \\
0348-0190850 &                 HD 300458 &   10:04:57.4 &  -55:10:37.5 &       F8 &     8.27 &     1.62 &        5700 &    -4.25 &                  ARG \\
0350-0195011 &           TYC 8603-1896-1 &   10:06:11.0 &  -54:58:29.7 &   K2$^*$ &     9.09 &     2.42 &        4950 &    -3.59 &                  CAR \\
0154-0097276 &                 CD-73 584 &   10:30:03.5 &  -74:32:32.2 &       K0 &     8.19 &     1.91 &        5300 &    -3.60 &              CAR/ARG \\
0349-0227761 &                         - &   10:40:10.6 &  -55:03:42.4 &   K6$^*$ &     9.45 &     3.40 &        4150 &    -2.99 &                  ARG \\
0277-0339323 &            TYC 8962-532-1 &   10:56:42.3 &  -62:17:41.0 &   K6$^*$ &     8.10 &     3.16 &        4250 &    -3.66 &          COL/CAR/ARG \\
0370-0341486 &            TYC 8617-909-1 &   11:23:47.0 &  -52:57:39.4 &   M0$^*$ &     7.95 &     3.71 &        3800 &    -3.49 &              ARG/COL \\
0104-0056948 &            TYC 9420-112-1 &   12:36:11.1 &  -79:31:38.4 &   K3$^*$ &     8.89 &     2.45 &        4650 &    -3.73 &              ARG/ABD \\
0568-0654078 &                         - &   17:06:22.7 &  -33:08:15.3 &   K7$^*$ &     8.62 &     3.54 &        4100 &    -3.39 &                  ARG \\
0405-0785226 &   2MASS J18151564-4927472 &   18:15:15.6 &  -49:27:47.3 &       M3 &     8.04 &     4.55 &        3450 &    -3.16 &             BPMG/ARG \\
0277-1048331 &                 HD 173167 &   18:48:06.4 &  -62:13:47.0 &      F5V &     6.14 &     1.16 &        6450 &    -4.35 &             BPMG/ARG \\
0601-0968580 &                         - &   18:58:04.7 &  -29:53:32.2 &   M0$^*$ &     8.76 &     4.08 &        3700 &    -2.93 &                 BPMG \\
0438-0823866 &   2MASS J19233820-4606316 &   19:23:38.2 &  -46:06:31.7 &       M0 &     8.27 &     3.77 &        3850 &    -3.21 &             BPMG/ARG \\
0577-1232093 &   2MASS J20055640-3216591 &   20:05:56.4 &  -32:16:59.2 &   M2$^*$ &     7.94 &     4.44 &        3550 &    -3.56 &                 BPMG \\
0706-0872415 &   2MASS J21100535-1919573 &   21:10:05.4 &  -19:19:57.6 &       M2 &     7.20 &     4.84 &        3350 &    -2.96 &                 BPMG \\
\hline                                                                          
\end{tabular}                                                                   
\end{center}                                                                    
\end{table*}

%%%%%%%%%%%%%%%%%%%%%%%%%%%%%%%%%%%%%%%%%%%%%%%%%%%%%%%%%%%%%%%%%%%%%%%%%%%%%%
                                                          
%%%%%%%%%%%%%%%%%%%%%%%%%%%%%%%%%%%%%%%%%%%%%%%%%%%%%%%%%%%%%%%%%%%%%%%%%%%%%%
%\input{table2.tex}
%\input{tables_v2/table2.tex}
\begin{table*}                                                                  
\setlength{\tabcolsep}{1.6mm}                                                   
\begin{center}                                                                  
\scriptsize                                                                     
\caption{Additional properties and membership status of the candidate          
members. (1) NOMAD-identifier. (2) Reference for source       
spectroscopic information. ANU and FEROS refer to our observations with the     
Echelle spectrograph at the 2.3-m telescope of the Australian National          
University and with the FEROS instrument at the 2.2\,m MPG/ESO                  
telescope, respectively. SACY: spectroscopic data taken from                    
\citet{torres2006}. (3) Heliocentric radial velocity.                           
(4) Measured lithium equivalent width. (5) Measured H$_{\alpha}$                
equivalent width {(if the line is in emission)}. (6) Spectroscopic binary candidates.             
(7) $\Delta M_{Ks}$ values (for definition see Sect.~\ref{selection})      
for all relevant kinematic groups listed in (12).            
(8) Derived $\chi_r^2$ values for all relevant kinematic groups.                
(9) Derived $d_{UVW}$ values for all relevant kinematic groups.                 
(10) Membership status based on the kinematic criteria described in             
Sect.~\ref{kinematiccrit}. '1' or '0'                                           
indicates whether the criterion is fulfilled                                    
or not.                                                                         
(11) Membership status based on                                                 
the lithium content. '1' or '0': the criterion is fulfilled or not,             
'-': lithium content cannot be used as a reliable                              
parameter {to judge the} membership status                                    
(see more in Sect.~\ref{lithiumcrit}) (12) Identifiers of candidate moving group(s).
\label{table2}}                                                                 
\begin{tabular}{ccrrrrrrrrrr}                                                   
\hline\hline                                                                    
    NOMAD ID &   Ref. &                 RV &   EW$_{\rm Li}$ &     EW$_{\rm H_{\alpha}}$ &    SB? &     $\Delta$M$_{Ks}$ &                $\chi_r^2$ &                 $d_{UVW}$ &   C$_{\rm kin}$ &    C$_{\rm Li}$ &          Candidate groups \\
             &        & [$\rm km\,s^{-1}$] &    [$\rm m\AA$] &               [$\rm \AA$] &        &                [mag] &                           &        [$\rm km\,s^{-1}$] &                 &                 &                           \\
         (1) &    (2) &                (3) &             (4) &                       (5) &    (6) &                  (7) &                       (8) &                       (9) &            (10) &            (11) &                      (12) \\
\hline
0367-0004557 &    ANU &     $+$9.4$\pm$0.8 &      105$\pm$12 &                         - &      - &    -0.18/-0.18/-0.03 &              1.3/7.3/17.1 &              2.1/5.6/11.1 &      1/0/0 &      1/1/1 &               COL/THA/ABD \\
0361-0008966 &    ANU &    $+$11.5$\pm$1.6 &       146$\pm$8 &                         - &      - &                -0.39 &                       0.7 &                       2.7 &          1 &          1 &                       COL \\
0540-0027893 &    ANU &    $+$18.8$\pm$0.8 &      240$\pm$14 &                         - &      - &                -0.13 &                      0.01 &                       0.3 &          1 &          1 &                       COL \\
0453-0032480 &    ANU &    $+$19.5$\pm$2.3 &      160$\pm$14 &                         - &      - &                +0.22 &                       0.8 &                       3.2 &          1 &          1 &                       COL \\
0454-0034540 &   SACY &    $+$19.3$\pm$0.2 &             285 &                         - &      - &                +0.45 &                       2.0 &                       3.6 &          1 &          1 &                       COL \\
0466-0044706 &    ANU &    $+$23.6$\pm$1.8 &      158$\pm$10 &                         - &      - &                +0.55 &                       0.3 &                       1.5 &          1 &          1 &                       COL \\
0466-0044732 &    ANU &    $+$22.6$\pm$6.7 &           $<$41 &                         - &      - &                +0.01 &                       0.2 &                       2.2 &          1 &         -- &                       COL \\
0211-0114754 &   SACY &    $+$18.7$\pm$1.0 &             226 &                         - &      - &                -0.20 &                       1.7 &                       3.5 &          1 &          1 &                       COL \\
0298-0039269 &    ANU &    $+$26.6$\pm$1.2 &       110$\pm$9 &                         - &      - &                -0.43 &                       5.2 &                       5.3 &          0 &          1 &                       COL \\
0211-0129478 &   SACY &    $+$17.5$\pm$0.5 &             265 &                         - &      - &                +0.00 &                       2.4 &                       3.7 &          1 &          1 &                       COL \\
0404-0048765 &  FEROS &    $+$23.3$\pm$0.5 &           $<$16 &                         - &      Y &                +0.10 &                       3.3 &                       4.5 &          1 &          0 &                       ARG \\
0351-0047432 &    ANU &    $+$22.5$\pm$1.0 &        76$\pm$9 &                         - &      - &                -0.07 &                       0.2 &                       1.2 &          1 &          0 &                       COL \\
0287-0048860 &    ANU &    $+$22.6$\pm$1.1 &        85$\pm$8 &                         - &      - &                -0.10 &                      0.06 &                       0.6 &          1 &          1 &                       CAR \\
0339-0066782 &  FEROS &    $+$26.1$\pm$2.6 &       39$\pm$10 &                         - &      Y &                +0.47 &                       0.5 &                       3.1 &          1 &          0 &                       COL \\
0315-0057897 &  FEROS &    $+$19.5$\pm$1.0 &      175$\pm$10 &                         - &      - &                +0.52 &                       2.7 &                       3.7 &          1 &          1 &                       CAR \\
0352-0070710 &  FEROS &    $+$19.8$\pm$0.4 &      222$\pm$10 &                         - &      - &                -0.04 &                       2.1 &                       3.5 &          1 &          1 &                       ARG \\
0368-0077090 &  FEROS &    $-$10.9$\pm$8.4 &           $<$26 &                         - &      Y &                -0.67 &                      10.9 &                      34.5 &          0 &          0 &                       COL \\
0263-0084896 &  FEROS &    $+$22.1$\pm$0.4 &      164$\pm$10 &                         - &      - &          +0.36/-0.04 &                  0.01/2.3 &                   0.3/3.0 &        1/1 &        1/1 &                   CAR/COL \\
0664-0176864 &  FEROS &    $+$24.7$\pm$1.1 &           $<$20 &                         - &      - &                +0.43 &                      0.01 &                       0.3 &          1 &          0 &                       COL \\
0331-0114466 &  FEROS &    $+$23.4$\pm$0.3 &      228$\pm$10 &                         - &      - &                +0.31 &                       0.4 &                       1.6 &          1 &          1 &                       CAR \\
0406-0105889 &  FEROS &    $+$24.3$\pm$0.3 &      219$\pm$11 &                         - &      - &                +0.29 &                       0.7 &                       1.8 &          1 &          1 &                       COL \\
0248-0110675 &  FEROS &    $+$21.7$\pm$0.4 &      257$\pm$10 &                         - &      - &          +0.27/-0.12 &                  0.03/1.4 &                   0.4/2.3 &        1/1 &        1/1 &                   CAR/COL \\
0248-0115043 &  FEROS &     $+$5.6$\pm$0.3 &      144$\pm$10 &                         - &      - &          -0.26/+0.67 &                  6.2/89.3 &                  5.9/14.6 &        0/0 &        1/1 &                   ARG/COL \\
0354-0106699 &  FEROS &    $+$14.6$\pm$0.4 &      164$\pm$10 &                         - &      - &                +0.41 &                      0.03 &                       0.5 &          1 &          1 &                       ARG \\
0341-0108561 &  FEROS &    $+$13.9$\pm$0.8 &      218$\pm$11 &          $-$0.20$\pm$0.06 &      - &                +0.33 &                       0.3 &                       2.4 &          1 &          1 &                       ARG \\
0489-0152365 &  FEROS &    $+$25.7$\pm$0.4 &      117$\pm$10 &                         - &      - &                +0.26 &                       2.4 &                       2.5 &          1 &          1 &                       COL \\
0269-0122275 &  FEROS &    $+$22.1$\pm$0.7 &      201$\pm$11 &                         - &      - &                -0.02 &                      0.10 &                       0.8 &          1 &          1 &                       CAR \\
0291-0099052 &  FEROS &    $+$22.3$\pm$0.4 &      246$\pm$10 &                         - &      - &                +0.49 &                       0.8 &                       2.6 &          1 &          1 &                       CAR \\
0372-0169783 &  FEROS &    $+$23.3$\pm$0.5 &      275$\pm$11 &          $-$0.38$\pm$0.10 &      - &          +0.39/+0.94 &                   1.0/0.3 &                   2.0/1.6 &        1/1 &        1/1 &                   COL/CAR \\
0313-0102916 &  FEROS &    $+$22.0$\pm$0.7 &      274$\pm$16 &                         - &      - &          -0.17/+0.29 &                   0.4/0.2 &                   1.2/1.3 &        1/1 &        1/1 &                   COL/CAR \\
0397-0143387 &   SACY &            $+$22.5 &             340 &                         - &      - &          +0.92/+0.36 &                   0.2/0.8 &                   1.9/3.0 &        1/1 &        1/1 &                   CAR/COL \\
0268-0125760 &  FEROS &    $-$10.3$\pm$1.8 &           $<$20 &                         - &      - &    +0.02/-0.91/+0.42 &            82.6/31.0/92.5 &            30.0/21.1/31.5 &      0/0/0 &      0/0/0 &               COL/ARG/CAR \\
0348-0138354 &  FEROS &    $+$33.8$\pm$3.9 &      252$\pm$14 &                         - &      Y &          +0.32/-0.17 &                   3.0/3.8 &                 11.6/13.1 &        0/0 &        1/1 &                   CAR/COL \\
0308-0116109 &  FEROS &    $+$21.6$\pm$1.1 &      276$\pm$15 &                         - &      Y &          +0.57/+0.13 &                  0.01/0.6 &                   0.3/2.2 &        1/1 &        1/1 &                   CAR/COL \\
0337-0139009 &  FEROS &    $+$20.0$\pm$0.3 &      107$\pm$10 &                         - &      - &          -0.20/+0.27 &                   0.2/2.1 &                   0.7/2.2 &        1/1 &        0/0 &                   COL/CAR \\
0415-0155444 &  FEROS &    $+$21.7$\pm$0.4 &       157$\pm$9 &                         - &      - &          +0.18/-0.37 &                   0.6/0.2 &                   1.4/1.3 &        1/1 &        1/1 &                   CAR/COL \\
0193-0211246 &  FEROS &    $+$19.4$\pm$0.4 &      213$\pm$10 &                         - &      - &          +0.09/-0.98 &                 0.02/14.5 &                  0.4/12.1 &        1/0 &        1/1 &                   CAR/ARG \\
0367-0179130 &  FEROS &    $+$22.6$\pm$0.8 &      315$\pm$10 &          $-$0.15$\pm$0.03 &      - &          +0.03/-0.45 &                   0.1/1.1 &                   0.7/2.1 &        1/1 &        1/1 &                   CAR/COL \\
0393-0168117 &  FEROS &     $+$5.8$\pm$0.4 &      277$\pm$10 &                         - &      - &          +0.24/+0.73 &                88.1/108.8 &                 15.0/16.4 &        0/0 &        1/1 &                   COL/CAR \\
0327-0202270 &  FEROS &    $+$20.4$\pm$2.3 &      171$\pm$13 &                         - &      Y &          -0.02/-0.44 &                  0.05/0.2 &                   0.9/1.5 &        1/1 &        1/1 &                   CAR/COL \\
0348-0190850 &  FEROS &    $+$11.8$\pm$0.8 &       37$\pm$10 &                         - &      - &                +0.23 &                       0.5 &                       2.0 &          1 &          0 &                       ARG \\
0350-0195011 &  FEROS &    $+$27.1$\pm$1.7 &       21$\pm$12 &                         - &      Y &                +0.17 &                       3.7 &                       6.4 &          0 &          0 &                       CAR \\
0154-0097276 &  FEROS &    $+$17.8$\pm$0.3 &      199$\pm$16 &                         - &      - &          +0.26/-0.64 &                 0.05/21.7 &                  0.6/12.7 &        1/0 &        1/1 &                   CAR/ARG \\
0349-0227761 &  FEROS &     $+$6.2$\pm$2.0 &           $<$35 &          $-$0.74$\pm$0.05 &      Y &                +0.63 &                       0.2 &                       1.8 &          1 &          0 &                       ARG \\
0277-0339323 &  FEROS &    $+$43.4$\pm$1.6 &       50$\pm$14 &          $-$1.23$\pm$0.10 &      - &    +0.51/+0.82/-0.27 &           79.9/73.6/113.6 &            27.2/25.1/37.2 &      0/0/0 &      0/0/1 &               COL/CAR/ARG \\
0370-0341486 &  FEROS &     $+$5.1$\pm$1.0 &       43$\pm$11 &          $-$0.97$\pm$0.12 &      - &          -0.02/+0.74 &                 0.04/19.3 &                  0.5/10.7 &        1/0 &        1/1 &                   ARG/COL \\
0104-0056948 &  FEROS &    $+$22.6$\pm$0.5 &      108$\pm$11 &                         - &      - &          +0.17/+0.65 &                  47.5/1.1 &                  20.8/3.1 &        0/1 &        0/1 &                   ARG/ABD \\
0568-0654078 &  FEROS &    $+$29.4$\pm$1.5 &           $<$27 &                         - &      - &                +0.46 &                     317.8 &                      49.5 &          0 &          0 &                       ARG \\
0405-0785226 &  FEROS &     $-$7.7$\pm$6.1 &           $<$46 &          $-$5.00$\pm$0.20 &      Y &          -0.34/-1.00 &                   0.4/1.5 &                   4.8/8.1 &        1/0 &         -- &                  BPMG/ARG \\
0277-1048331 &  FEROS &     $+$0.8$\pm$7.0 &      107$\pm$16 &                         - &      Y &          +0.15/-0.97 &                  0.03/2.1 &                  1.1/11.0 &        1/0 &        1/1 &                  BPMG/ARG \\
0601-0968580 &  FEROS &     $-$7.0$\pm$0.3 &           $<$28 &          $-$4.90$\pm$0.20 &      - &                +0.08 &                       0.3 &                       2.2 &          1 &         -- &                      BPMG \\
0438-0823866 &  FEROS &     $-$0.2$\pm$1.2 &      422$\pm$12 &          $-$1.13$\pm$0.15 &      - &          +0.04/-0.66 &                  0.7/46.4 &                  3.1/15.7 &        1/0 &        1/1 &                  BPMG/ARG \\
0577-1232093 &  FEROS &     $-$5.1$\pm$1.3 &      140$\pm$16 &          $-$1.92$\pm$0.15 &      - &                -0.09 &                       0.4 &                       1.7 &          1 &          1 &                      BPMG \\
0706-0872415 &  FEROS &     $-$6.3$\pm$1.2 &           $<$41 &          $-$3.21$\pm$0.15 &      - &                -0.12 &                       0.3 &                       1.3 &          1 &         -- &                      BPMG \\
\hline                                                                          
\end{tabular}                                                                   
\end{center}                                                                    
\end{table*}

%%%%%%%%%%%%%%%%%%%%%%%%%%%%%%%%%%%%%%%%%%%%%%%%%%%%%%%%%%%%%%%%%%%%%%%%%%%%%%

\section{Observations and data reduction} \label{obsanddatared}

\subsection{Spectroscopic observations}

Our sample includes 54 candidates. For four of them, spectroscopic data were taken from the SACY 
catalogue \citep{torres2006}. 
For the remaining 50 selected candidate stars we obtained high-resolution spectra using two different instruments.

Nine spectra were obtained in 2009
July and August with the Echelle spectrograph
at the 2.3-m telescope of the Australian National University
(Siding Spring Observatory, Australia).
The spectra covered the whole visual range in 27 {E}chelle orders between 
3900 \AA\ and 6720 \AA\, with only small gaps between the three reddest orders. 
The nominal spectral resolution is $\lambda/\Delta \lambda\approx$ 
23~000 at the H$\alpha$ line (6563 \AA).
The integration time per object ranged from 180\,s to 1800\,s,
depending on the brightness, and we achieved 
signal-to-noise ratios of about 100 around the H$\alpha$ line.
All data were reduced with standard 
IRAF\footnote{IRAF is distributed by the National Optical Astronomy 
Observatories, which are operated by the
Association of Universities for Research in Astronomy, Inc., 
under cooperative  agreement with the National Science Foundation.} tasks, 
including bias and flat-field corrections, cosmic ray removal, 
extraction of the 27 individual orders of the echelle spectra, 
wavelength calibration, and continuum normalization. ThAr spectral lamp 
exposures were regularly taken before and after every object spectrum 
to monitor the wavelength shifts
of  the spectra on the CCD.

41 additional targets were observed
using the FEROS instrument 
\citep[the Fiber-fed Extended Range Optical Spectrograph,][]{kaufer1999} 
mounted at the 2.2\,m MPG/ESO 
%(Max-Planck-Gesellschaft/European Southern Observatory) 
telescope 
in La Silla, Chile, between 2011 April 15 and 18.
FEROS is an {\'E}chelle spectrograph that covers the spectral region between 
3500 and 9200 \AA\ in 39 echelle orders in one exposure, providing a 
spectral resolution 
of $R\approx$48~000. We used the "object-sky" mode in our observations 
in which one of the two fibers is positioned at the target while the 
other one observes the sky. The integration times ranged between 180\,s and 
3200\,s depending on the target brightness. The resulting 
signal-to-noise {ratio} was $\gtrsim$60 at the lithium line (Li\,I, 6708 \AA).
Standard calibration observations, including bias, dark, and flat exposures, 
as well as Th-Ar lamp calibration spectra, were obtained at each night.
The data were processed using the FEROS data reduction system (DRS) provided 
by ESO.
The reduction procedure includes bias subtraction, flat-field 
correction, 
background subtraction, the definition and extraction of orders, and  
wavelength calibration.

\subsection{Obtaining the radial velocities (RVs)}

All spectra were evaluated according to the standard fashion in IRAF.
Template spectra were taken from \citet{munari2005}. 
%While fitting the velocities, we selected those models that 
%have the closest effective temperature to the one 
%derived from the photometry, and 
%for which [Fe/H]=0, $v\sin i=0$, $\log g=4.5$. 
{In the velocity fitting procedure we selected those 
models that provided the closest effective temperature to that derived 
from the photometry and that had [Fe/H]=0, $v\sin i=0$, $\log g=4.5$.} 
Cross correlation functions (CCFs) were calculated by the {\sc fxcor} task in
IRAF, in the 5200-6400 K range. Continuum normalization is a difficult
task in this spectral range, especially for late-type stars,
mostly because of {the presence of} molecular bands. Both observed and model spectra
were normalized to a 13th order polinomial model. This model likely
clipped the widest bands, too, but resulted in a satisfactorily smooth
correlation function, unaffected by {the misestimation of the band shapes}. 
%how models misestimated the shape
%of the bands.

During the whole task, the most challenging problem was the crude
temperature resolution of model spectrum libraries, and in many cases,
template and observed spectra exhibited prominent differences in the
continuum/molecular band level. We tested how the misestimation of the template
spectra affects the radial velocity determination. The test led to
statisfactory results: although the calculated {\it error} of the
RV increased a lot with non-adequate template spectra,
the estimated RV value did not change significantly. Therefore we are
confident {of} the reliability of the fitted RVs,
{however,} we have to note that a considerable part of the errors may have
come from the lack of a template spectrum with appropriate
temperature. The resulting RVs are listed in Table~\ref{table2}.

\subsection{Identification of spectroscopic binaries} \label{cbs}

Candidate spectroscopic binaries were identified by
complex structures in the CCF, in most cases two maxima {could be found with} 
similar heights. In
some cases, the second maximum was not observed, but a prominent
skewness of the CCF, or a plateau on the wings appeared, likely due to
a companion. {For} these targets the CCF were fitted manually assuming two
stars, and we derived the velocities and {the velocity differences}. (In this
paper, we use the velocity derived for the brighter component, if binarity was suspected.)

The calculated velocity differences were typically in the
20--50~km~s$^{-1}$ range, which is compatible with orbital periods of
several 10 days to a year, although this is not an
exact estimate because we only have velocity amplitudes from one
measurement. In the case of 0368-0077090, the binarity was confirmed by the
variation of the radial velocity of the brighter component, as
observations taken at different epochs led to radial velocities that
were incompatible with a constant value.

\subsection{Measuring the equivalent width of the Li~I and H$_{\alpha}$ lines} 

We used the IRAF {\sl continuum} task with a cubic spline algorithm to
model the local continuum in the wavelength ranges between 6680 and 6736~\AA\ 
and between 6520 and 6605~\AA\
around the lithium and H$_{\alpha}$ lines, respectively,
normalizing all fluxes to unity.
The equivalent widths of lithium absorption lines (EW$_{\rm Li}$) 
were measured by fitting Gaussian profiles using a self-developed IDL routine. 
A measurable amount of Li has been detected in the spectra
 of 40 stars. For the rest of the sample (10 targets),
 we estimated upper limits for the equivalent widths
 using our {code}. 
Equivalent widths of the H$_{\alpha}$ line were also 
obtained for stars where this line was in emission.
%in the wavelength range between 6680 and 
%6736 \AA\ in the case of lithium absorption line 
The measured Li\,I and H$_\alpha$ equivalent widths are listed in 
Table~\ref{table2}.

\section{Results} \label{results}

\subsection{Assignment decision}

\begin{figure} 
\includegraphics[width=85mm]{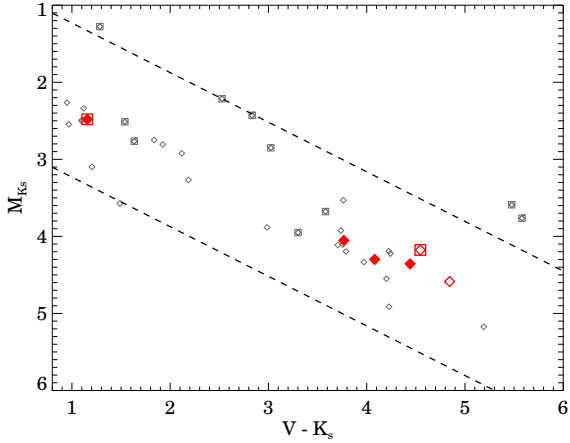}
\caption{Absolute K$_{\rm s}$ magnitude versus V$-$K$_{\rm s}$ 
colour diagram for the known and candidate members of BPMG. 
Known members are displayed by small grey diamonds.
Larger black (red in the electronic version) diamonds 
represent our candidate stars taken from Table~\ref{stellarprop}.  
Squares denote those close binaries for which no photometric 
data are available separately for the individual components.
Following the outline of \citet{lepine2009} 
the loci of the known members of the specific groups
were fitted with a linear, the $\pm$1 magnitude range of the fit is 
represented by dashed lines.
Nearly all known members are located within these lines in the diagram.
Candidate objects {outside this} area were omitted from our survey 
(see Sect.~\ref{selection}).
Those candidates that are finally classified as likely members of BPMG 
are marked by filled symbols.
}
\label{cmd}
\end{figure}

\begin{figure} 
\includegraphics[width=85mm]{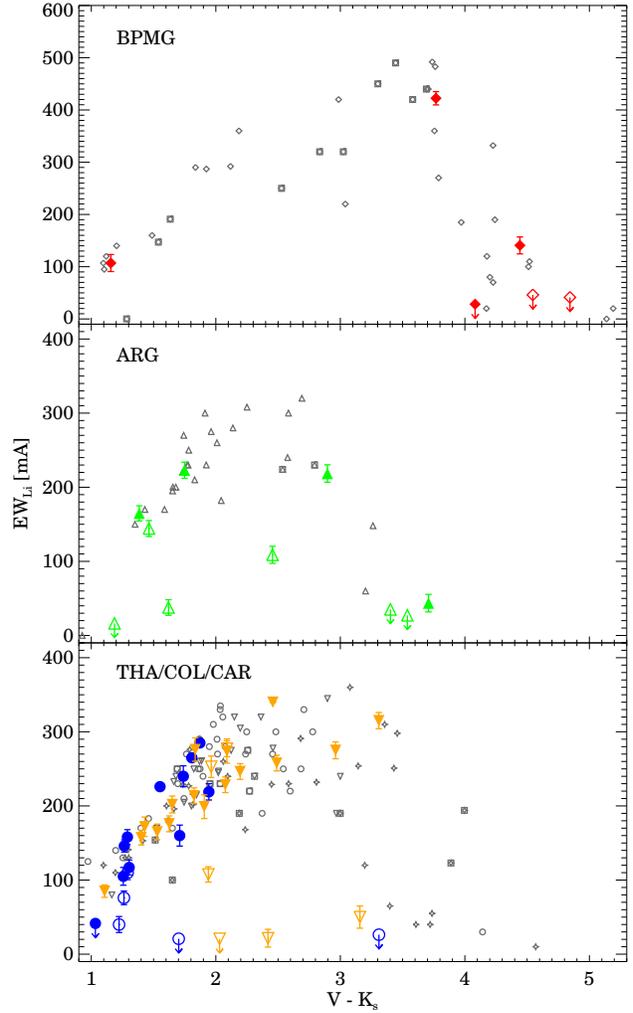}
\caption{Equivalent width of lithium as a function of V$-$K$_{\rm s}$.
Candidate members of BPMG, ARG, COL and CAR are {marked} by 
diamonds (red in the electronic version), 
triangles (green in the electronic version), 
circles (blue in the electronic version)  
and upside-down triangles (orange in the electronic version), respectively.
Known members of a specific group are represented by similar but smaller grey 
symbols than the candidates. Known THA members are displayed by small grey 
star symbols. 
Candidates that turned out to be likely members of a specific group 
are marked by filled symbols.
}
\label{ewli}
\end{figure}

%\input{membership.tex}
%%%%%%%%%%%%%%%%%%%%%%%%%%%%%%%%%%%%%%%%%%%%%%%%%%%%%%%%%%%%%%%%%%%%%%%%%%%%
Members of young moving groups can be identified based on their common 
space motion within the Galaxy and their common ages.
Supplementing astrometric data by new radial velocity measurements, the  
similarity between the Galactic space motion of the 
candidate stars and characteristic space motion of 
groups can be evaluated more reliably than based only 
on proper motion data (especially in those cases where 
measured trigonometric parallaxes are also available).   
As a preliminary age diagnostic 
in the sample selection, we kept only those candidates 
that exhibit{ed} coronal activity similar to the known member of 
young moving groups and whose position in the colour-magnitude diagram 
coincide{d} well with the loci of the known members (Fig.~\ref{cmd}). 
By comparing the measured Li contents with the Li distribution
of the specific association(s) we can investigate further whether 
our targets could be co-eval with the appropriate group.

%\input{kinematiccrit.tex}
%%%%%%%%%%%%%%%%%%%%%%%%%%%%%%%%%%%%%%%%%%%%%%%%%%%%%%%%%%%%%%%%%%%%%%%%%%%%%
\subsubsection{Kinematic criteria} \label{kinematiccrit}

Using the new radial velocity data we computed the Galactic space velocity for 
each candidate star. 
We used a right-handed UVW 
coordinate system (U is positive towards the Galactic centre, 
V is positive in the direction of galactic rotation and W is positive 
towards the North Galactic pole) and followed the general recipe described 
in "The Hipparcos and Tycho Catalogues"(ESA 1997). 
Coordinates and proper motions of stars were taken from the NOMAD 
catalogue. For targets included in the Hipparcos catalogue 
we used their measured trigonometric parallaxes, otherwise 
 we adopted the derived kinematic parallaxes. Radial velocity data were taken 
 from Table~\ref{table2}.    
For 0466-0044732 (HIP 25436) and 0466-0044706 (HIP 25434), which 
 constitute a binary system according to the 
 {\sl The Washington Visual Double Star Catalog} \citep{mason}, 
 we used the weighted average of their measured trigonometric 
 parallaxes. 	

In order to characterize the kinematic match between the candidates 
and different groups, we followed \citet{shkolnik2012} by 
defining two parameters: 
(1) a velocity modulus 
$d_{UVW} = \sqrt{ (U-U_0)^2 + (V-V_0)^2 + (W-W_0)^2) }$
that shows the distance of the measured $U, V, W$ 
components to the characteristic space motion of the different 
moving groups ($U_0, V_0, W_0$), and (2) a
reduced $\chi^2$ statistics with 3 degrees of freedom, 
$\chi_r^2 = \frac{1}{3}\left[\frac{(U - U_0)^2} {({\sigma_{U}}^2 + 
         {\sigma_{U_0}}^2)} + 
          \frac{(V - V_0)^2} {({\sigma_{V}}^2 + {\sigma_{V_0}}^2)} +
 \frac{(W - W_0)^2} {({\sigma_{W}}^2 + {\sigma_{W_0}}^2)} \right]$,
where $\sigma_{U}, \sigma_{V}, \sigma_{W}$ are the uncertainties 
of the measured $U, V, W$ velocities of the star, 
 while $\sigma_{U_0}, \sigma_{V_0}, \sigma_{W_0}$ 
are the uncertainties of characteristic space motion of 
the specific moving group \citep[taken from][]{torres2008}. 
We computed and listed these values in Table~\ref{table2} 
for each star with respect to each relevant group.  
We also checked the resulting $d_{UVW}$ and $\chi_r^2$ values 
for objects already assigned to a group in the literature.
In accordance with \citet{shkolnik2012},
for most of the known members ($\gtrsim$90\%) we obtained 
$d_{UVW}<5$~km~s$^{-1}$ and $\chi_r^2<6$, therefore 
we set these values as a lower limit of membership
for the new candidates as well. 
Membership status of our targets, derived from
these kinematic criteria (C$_{kin}$), is reviewed in Table~\ref{table2}.   
We note that in some cases the target 
could be assigned to several different moving groups but the new radial velocity data 
helped to constrain the assignment for one specific group (e.g. 0367-0004557).
 
In the case of candidate spectroscopic binaries with 
double peak in the correlation function we included the RV of the
brighter peak in the analysis (see Sect.~\ref{cbs}). 
Therefore, RV values of these systems are biased
estimates of the systemic velocity, with currently unknown bias values that may reach even
10 km~s$^{-1}$ in some cases. There are also candidate spectroscopic binaries 
where the correlation function had significant skewness, but exhibited only a single prominent peak, which 
gives also a biased estimate of the systemic velocity. 
 The proper disentangling of the components will
require higher resolution, higher signal-to-noise ratio, and multi-epoch observations. This approach
%may lead to the rejection of real group members or the acceptance of false group
%members, and the proper handling of the problem will require further investigations.
makes the assignment of the specific objects less reliable and can lead to
the rejection/acceptance of a real/false group member. 

%\input{lithiumcrit.tex}
%%%%%%%%%%%%%%%%%%%%%%%%%%%%%%%%%%%%%%%%%%%%%%%%%%%%%%%%%%%%%%%%%%%%%%%%%%%%%
\subsubsection{Membership status based on lithium content} \label{lithiumcrit}

Due to the fact that lithium is destroyed at relatively low temperatures 
(2.5$\times$10$^6$\,K) in stellar interiors, the primordial lithium 
depletes with time in stars with a convective layer, {that makes this} 
element one of the best age indicators for young stars \citep{zucksong2004}.
We used the equivalent width
of the lithium line (EW$_{\rm Li}$) as an empirical measurement of the lithium
abundances in our targets. In accordance with our general 
principle, 
for membership of a candidate we requested that its 
lithium content {is} consistent with those of known members.
In Fig.~\ref{ewli} 
we displayed the measured EW$_{\rm Li}$ values 
of our candidates and {that of} known members of specific groups,
as a function of $V-K_s$ colour indices. 
Members and member candidates of THA, COL, and CAR groups 
are plotted together since from the Li abundance point of view these three
groups are practically indistinguishable \citep{daSilva2009}.
In order to evaluate whether the Li content of a 
certain candidate is consistent with the Li distribution of 
the assigned group, we compared the target's measured EW$_{\rm Li}$
to that of known members {with} similar $V-K_s$ indices. 
For each target we selected those known members of the assigned group
for comparison whose $V-K_s$ indices are within a $\pm r_{V-K_s}$ magnitude 
range of the candidates star's $(V-K_s)_c$, $r_{V-K_s}$ was set to 0.07~mag if 
$(V-K_s)_c<3$~mag and 0.1~mag if $(V-K_s)_c\geq3$~mag. 
A candidate was accepted as a member -- in terms of lithium abundance -- 
if its EW$_{\rm Li}\pm \sigma_{EW_{Li}}$  value 
falls in the EW$_{\rm Li}$ range of the selected known members.
For some ARG candidate stars where we could not find any known members within the abovementioned 
colour index range, we used GAYA (THA/COL/CAR) stars -- whose age is close to that of ARG --
for this comparison. 
The rate of lithium depletion depends also on the effective stellar temperature. 
In early M-type stars the Li can be entirely depleted 
by an age of $\sim$10~Myr \citep{zucksong2004,mentuch2008}, 
thus the lack of lithium in the spectrum of such stars cannot be used to unambigously
reject the membership in groups with ages $>$10~Myr. 
 
 Membership status of our targets, derived based on 
this criterion (C$_{Li}$), is listed in Table~\ref{table2}.     

%\input{finalmembership.tex}
%%%%%%%%%%%%%%%%%%%%%%%%%%%%%%%%%%%%%%%%%%%%%%%%%%%%%%%%%%%%%%%%%%%%%%%%%%%%%
\subsubsection{Final membership}  \label{finalassignment}

For the final membership we demanded that both our kinematic and 
lithium content based criteria {are} fulfilled (C$_{kin}$ and C$_{Li}$ = 1).
This means that taking into account our preliminary selection criteria, 
our targets have comparable X-ray luminosity and lithium abundance 
than those of the known members
with similar colour indices, their position in the colour-magnitude 
diagram are consistent with the loci of known members, and their {G}alactic 
space motion and location are also consistent with the characteristic 
space motion and position of the assigned group.    
According to our classification, two   
M-type stars (0405-0785226 and 0706-0872415) for which the lithium-based
diagnostic method cannot be used to support or reject the membership 
status were not adopted as probable new members of BPMG. 
0466-0044732 and 0601-0968580,
 whose lithium content was also not relevant 
in the classification, were approved as probable new members because 
they are located in binary systems where the other companion 
is considered as probable member of the specific group (see Sect.~\ref{notes}).
There are three candidate binary systems (0308-0116109, 0327-0202270, and 0277-1048331) 
{which fulfilled} both the kinematic and lithium content based criteria. 
Since RV values of these systems can only be considered as biased estimates 
of the systemic radial velocity, their U, V, W values are less reliable and 
further observations are needed to finally consolidate our assigments. 
Therefore, 0308-0116109 and 0327-0202270 were labeled as possible members 
only. Despite the binarity issue, 0277-1048331 was classified as a probable 
member because it has a widely
separated co-moving pair (TYC~9073-0762, for details see Sect.~\ref{notes}) 
that is found to be a member of 
the BPMG by \citet{torres2008}. 
As a result of our survey, we identified 33 new probable and 2 new possible members (see in Table~\ref{table3})
of four young nearby moving groups (BPMG, COL, CAR, ARG). 
All probable members have $d_{UVW} < 3.73$~{km~s$^{-1}$}, while the largest $\chi_r^2$ is 
 2.69. Assuming that the derived $\chi_r^2$ values follow a ${\chi }^2$ distribution 
with three degrees of freedom \citep[see also][]{shkolnik2012} the kinematic membership probability 
corresponding to $\chi_r^2 = 2.69$ is 0.44. Thus, all 33 selected stars exhibit 
a membership probability of at least 44\%, a strong indication of membership.   

There are several targets in our candidate list that can be assigned to more 
than one kinematic groups. In some cases, our new RV and Li content-based analysis 
helped to resolve this ambiguity 
leaving only one potential group (e.g. 0367-0004557). 
In eight cases (0263-0084896, 0248-0110675, 0372-0169783, 0313-0102916, 
0397-0143387, 0308-0116109, 0415-0155444, 0367-0179130), however, the 
ambiguity remained. These objects can either 
be assigned to the Columba or the Carina moving groups. These two groups have very similar 
space motion and somewhat overlapping spatial location. 
In order to establish the more probable group for these 
 targets, we compared 1) their distances to the mean position 
of the COL and CAR groups \citep[$d_{XYZ}$, mean X, Y, Z positions were 
taken from,][]{torres2008} and 2) their $d_{UVW}$ and $\chi_r^2$ values were
computed for the two relevant groups. 
For seven ambiguous targets (0263-0084896, 0248-0110675, 0372-0169783, 0313-0102916, 
0397-0143387, 0308-0116109, 0367-0179130) the assigment to 
the Carina association 
is more likely based on these parameters, because
$d^{CAR}_{UVW} \leq d^{COL}_{UVW}$, $\chi^2_{r,CAR} \leq \chi^2_{r,COL}$ and 
$d^{CAR}_{XYZ} < d^{COL}_{XYZ}$. The case of 0415-0155444 is somewhat more difficult 
because its space motion matches the values of COL ($d^{COL}_{UVW} = 1.26$~km~s$^{-1}$, 
$\chi^2_{r,COL} = 0.24$ while $d^{CAR}_{UVW} = 1.38$~km~s$^{-1}$, $\chi^2_{r,CAR} = 0.6$) better, however, the star 
is located significantly closer to the mean position of CAR ($d^{CAR}_{XYZ} = $22.2\,pc, while 
$d^{COL}_{XYZ} = $93.0\,pc). This star was finally assigned also to the Carina group 
due to the better match in the space position. 

Trigonometric parallaxes measured by {\sl Hipparcos} are available only for 
4 stars. These targets can be considered as new secure young moving group 
members.  For the remaining 31 stars, we applied kinematic distances 
computed based on {the} assumption that they belong to the specific kinematic group.
By deriving kinematic distances for the abovementioned four secure members, as well as 
for all those previously reported members of the selected groups for which
{\sl Hipparcos} data were also available \citep[taken from][]{torres2008} we found that 
the kinematic and trigonometric parallaxes agree within the 2$\sigma$ uncertainties 
for nearly all targets (and within 1$\sigma$ for 75\% of the sample).
Despite this good agreement and {also}
that kinematic and photometric distances of the 31 stars are consistent, 
 we cannot completely exclude that their 
real distance in fact deviates from the predicted one, thus their kinematic properties do 
not match those of the moving group. 
Based on the present data, these 31 stars can be 
classified as probable (29 cases) or possible (2 objects) new members, 
however, these objects yet lack trigonometric 
parallaxes, and further observations
(e.g. by the {\sl Gaia} mission) are necessary to prove our 
membership assignment.

In summary, we identified 4 secure members of the COL group, and suggested 
31 further probable/possible members of various nearby groups. 
%Of these, 4-4 stars 
%Four stars can be assigned to both the BPMG and the ARG groups, 
%while 11 and 16 stars may be members of the COL and CAR groups, respectively.
{We assign 4 stars to the BPMG, 4 stars to the ARG group, 11 stars to COL, and 16 stars
to CAR.}

Interestingly, our observations revealed that the measured parameters of 0104-0056948 
fulfill all the criteria of ABD membership, meaning that we serendipitiously identified 
a new probable AB Doradus member as well. Kinematic properties of this star were also 
listed in Table~\ref{table3}.

There are four stars in our sample 
(0454-0034540, 0211-0114754, 0211-0129478, 0397-0143387)
whose spectroscopic data were taken from the SACY 
survey \citep{torres2008}. In our analysis we concluded that 
these stars are likely members of the Columba/Carina groups, while  
\citet{torres2008} rejected them as a member of any of the five groups studied in our work. 
The different conclusions about group membership can be explained 
by the difference of the applied candidate selection algorithm and -- in the case of 0397-0143387 -- 
by the difference of the utilized astrometric data. Namely, for 0454-0034540 the $d_{UVW}$ value of 3.6~kms$^{-1}$ 
is one of the highest among our sample. \citet{torres2008} may have rejected this object 
due to their more strict kinematic criterion 
($\sqrt{ (U-U_0)^2 + (V-V_0)^2 + (W-W_0)^2 } < 3.5$~kms$^{-1}$). 
The X space coordinate of 0211-0114754 and 0211-0129478 is out of the boundaries of the Columba 
group specified by \citet[][in their Table 2]{torres2008}. In the case of 0397-0143387, 
the likely erroneous {TYCHO}-based proper motions may have led to the rejection 
of the candidate (for more details, see Sect.~\ref{notes}).

%\input{tables_v2/table3.tex}
%%%%%%%%%%%%%%%%%%%%%%%%%%%%%%%%%%%%%%%%%%%%%%%%%%%%%%%%%%%%%%%%%%%%%%%%%%%%%
                                      
\begin{table*}                                                                  
\setlength{\tabcolsep}{1.6mm}                                                   
\begin{center}                                                                  
\scriptsize                                                                     
\caption{Final list of probable and possible new members. (1) NOMAD-identifier.        
(2) Parallax. Asterisks indicate parallaxes taken from                          
\citet{vanleeuwen07}, otherwise kinematic parallaxes (see Sect.~\ref{selection})
are listed. For 0466-0044732 and 0466-0044706 that                             
 {form} a binary system (see Sect.~\ref{kinematiccrit})            
 we listed the weighted average of their measured trigonometric                 
 parallaxes.                                                                    
(3) $\Delta M_{Ks}$ values.                                               
(4)-(6) Galactic space velocity components (U,V,W)                              
of the star. (7)-(8) Derived $\chi_r^2$ and                                     
$d_{UVW}$ values for the assigned moving group.                                 
For a detailed definition of these parameters see Sect.~\ref{kinematiccrit}.    
(9) Multiplicity.                                                               
(10) Identifier                                                                 
of the assigned moving group. Asterisks indicate those candidate spectroscopic 
binary systems that are classified only as possible members in our work 
(see Sect.~\ref{finalassignment} for details).                                                    
\label{table3}}                                                                 
\begin{tabular}{crrrrrrrcc}                                                     
\hline\hline                                                                    
    NOMAD ID &              $\pi$ &   $\Delta$M$_{Ks}$ &                     U &                     V &                     W &            $\chi_r^2$ &             $d_{UVW}$ &   Multiplicity &      Group \\
             &              [mas] &              [mag] &    [$\rm km\,s^{-1}$] &    [$\rm km\,s^{-1}$] &    [$\rm km\,s^{-1}$] &                       &    [$\rm km\,s^{-1}$] &                &            \\
         (1) &                (2) &                (3) &                   (4) &                   (5) &                   (6) &                   (7) &                   (8) &            (9) &       (10) \\
\hline
0367-0004557 & 22.30$\pm$0.60$^*$ &              -0.18 &       $-$13.3$\pm$0.5 &       $-$23.7$\pm$0.6 &        $-$5.0$\pm$0.7 &      1.32 &      2.12 &              N &        COL \\
0361-0008966 & 10.17$\pm$0.85$^*$ &              -0.39 &       $-$12.3$\pm$1.3 &       $-$23.8$\pm$1.8 &        $-$4.3$\pm$1.5 &      0.65 &      2.70 &              N &        COL \\
0540-0027893 &   10.06$\pm$0.72~~ &              -0.13 &       $-$13.1$\pm$0.9 &       $-$21.6$\pm$1.2 &        $-$5.6$\pm$1.1 &      0.01 &      0.31 &              Y &        COL \\
0453-0032480 &    9.88$\pm$0.81~~ &               0.22 &       $-$10.5$\pm$1.5 &       $-$22.1$\pm$1.8 &        $-$4.2$\pm$1.9 &      0.77 &      3.17 &              N &        COL \\
0454-0034540 &   11.03$\pm$0.94~~ &               0.45 &       $-$10.5$\pm$0.7 &       $-$21.3$\pm$0.8 &        $-$3.6$\pm$0.9 &      2.01 &      3.60 &              N &        COL \\
0466-0044706 & 14.09$\pm$1.40$^*$ &               0.55 &       $-$11.7$\pm$0.9 &       $-$21.7$\pm$1.5 &        $-$5.8$\pm$1.3 &      0.31 &      1.53 &              Y &        COL \\
0466-0044732 & 14.09$\pm$1.40$^*$ &               0.01 &       $-$11.6$\pm$2.2 &       $-$20.3$\pm$5.3 &        $-$5.9$\pm$3.8 &      0.16 &      2.19 &              Y &        COL \\
0211-0114754 &    9.65$\pm$0.64~~ &              -0.20 &       $-$11.7$\pm$1.1 &       $-$22.5$\pm$1.0 &        $-$2.9$\pm$0.9 &      1.73 &      3.46 &              N &        COL \\
0211-0129478 &    9.68$\pm$0.65~~ &               0.00 &       $-$12.1$\pm$1.2 &       $-$21.3$\pm$0.7 &        $-$2.4$\pm$0.7 &      2.35 &      3.73 &              N &        COL \\
0287-0048860 &   12.46$\pm$1.24~~ &              -0.10 &        $-$9.9$\pm$1.1 &       $-$23.4$\pm$1.1 &        $-$4.2$\pm$0.9 &      0.06 &      0.56 &              N &        CAR \\
0315-0057897 &   11.51$\pm$1.42~~ &               0.52 &       $-$10.0$\pm$1.3 &       $-$19.7$\pm$1.0 &        $-$2.7$\pm$0.7 &      2.69 &      3.73 &              N &        CAR \\
0352-0070710 &    8.50$\pm$0.67~~ &              -0.04 &       $-$22.2$\pm$2.0 &       $-$17.9$\pm$0.5 &        $-$5.2$\pm$1.0 &      2.10 &      3.47 &              N &        ARG \\
0263-0084896 &   10.69$\pm$1.04~~ &               0.36 &       $-$10.1$\pm$1.3 &       $-$23.1$\pm$0.4 &        $-$4.1$\pm$0.5 &      0.01 &      0.29 &              N &        CAR \\
0331-0114466 &    9.69$\pm$1.32~~ &               0.31 &       $-$10.0$\pm$1.6 &       $-$23.6$\pm$0.3 &        $-$2.9$\pm$0.8 &      0.42 &      1.65 &              N &        CAR \\
0406-0105889 &    8.19$\pm$1.02~~ &               0.29 &       $-$13.6$\pm$1.7 &       $-$22.8$\pm$0.4 &        $-$4.4$\pm$1.1 &      0.71 &      1.82 &              N &        COL \\
0248-0110675 &    9.98$\pm$1.04~~ &               0.27 &       $-$10.2$\pm$1.6 &       $-$23.1$\pm$0.5 &        $-$4.8$\pm$0.7 &      0.03 &      0.44 &              N &        CAR \\
0354-0106699 &    8.37$\pm$0.67~~ &               0.41 &       $-$22.0$\pm$2.2 &       $-$14.8$\pm$0.4 &        $-$4.7$\pm$1.4 &      0.03 &      0.50 &              N &        ARG \\
0341-0108561 &    9.44$\pm$0.99~~ &               0.33 &       $-$21.6$\pm$2.9 &       $-$14.1$\pm$0.8 &        $-$7.4$\pm$2.1 &      0.31 &      2.41 &              N &        ARG \\
0489-0152365 &    8.39$\pm$0.97~~ &               0.26 &       $-$13.1$\pm$1.3 &       $-$24.2$\pm$0.5 &        $-$6.6$\pm$1.1 &      2.40 &      2.54 &              N &        COL \\
0269-0122275 &    8.09$\pm$1.07~~ &              -0.02 &       $-$10.1$\pm$2.3 &       $-$23.5$\pm$0.8 &        $-$5.1$\pm$1.6 &      0.10 &      0.84 &              N &        CAR \\
0291-0099052 &    9.20$\pm$1.21~~ &               0.49 &       $-$10.0$\pm$2.2 &       $-$24.0$\pm$0.6 &        $-$2.0$\pm$1.5 &      0.79 &      2.62 &              N &        CAR \\
%0372-0169783 &    9.43$\pm$1.04~~ &               0.39 &       $-$12.8$\pm$2.1 &       $-$23.3$\pm$0.5 &        $-$7.1$\pm$1.5 &      0.98 &      2.02 &              N &        CAR \\
0372-0169783 &   12.14$\pm$1.49~~ &               0.94 &        $-$9.7$\pm$1.7 &       $-$23.3$\pm$0.5 &        $-$5.9$\pm$1.2 &      0.27 &      1.63 &              N &        CAR \\
0313-0102916 &    8.52$\pm$1.01~~ &               0.29 &       $-$10.0$\pm$1.9 &       $-$22.7$\pm$0.7 &        $-$5.6$\pm$1.1 &      0.16 &      1.25 &              N &        CAR \\
0397-0143387 &   10.28$\pm$2.08~~ &               0.92 &       $-$10.7$\pm$3.1 &       $-$22.7$\pm$0.1 &        $-$2.5$\pm$2.1 &      0.23 &      1.95 &              N &        CAR \\
0308-0116109 &    8.63$\pm$1.08~~ &               0.57 &       $-$10.2$\pm$2.1 &       $-$23.0$\pm$1.1 &        $-$4.1$\pm$1.3 &      0.01 &      0.30 &              Y &        CAR$^*$ \\
0415-0155444 &   10.55$\pm$0.97~~ &               0.18 &       $-$10.5$\pm$1.1 &       $-$21.9$\pm$0.4 &        $-$3.6$\pm$0.6 &      0.60 &      1.38 &              N &        CAR \\
0193-0211246 &    9.34$\pm$0.93~~ &               0.09 &       $-$10.1$\pm$1.9 &       $-$23.2$\pm$0.9 &        $-$4.1$\pm$1.2 &      0.02 &      0.35 &              N &        CAR \\
0367-0179130 &    8.34$\pm$1.35~~ &               0.03 &       $-$10.1$\pm$2.7 &       $-$23.6$\pm$0.8 &        $-$4.6$\pm$1.9 &      0.10 &      0.68 &              N &        CAR \\
0327-0202270 &    8.67$\pm$0.89~~ &              -0.02 &       $-$10.4$\pm$1.8 &       $-$22.2$\pm$2.3 &        $-$4.1$\pm$1.2 &      0.05 &      0.91 &              Y &        CAR$^*$ \\
0154-0097276 &   12.26$\pm$0.89~~ &               0.26 &       $-$10.1$\pm$1.4 &       $-$23.1$\pm$0.7 &        $-$5.0$\pm$0.6 &      0.05 &      0.63 &              N &        CAR \\
0370-0341486 &   21.19$\pm$1.30~~ &              -0.02 &       $-$22.2$\pm$1.7 &       $-$14.1$\pm$1.2 &        $-$4.6$\pm$0.6 &      0.04 &      0.53 &              N &        ARG \\
0104-0056948 &   13.22$\pm$0.99~~ &               0.65 &        $-$6.9$\pm$1.6 &       $-$28.6$\pm$1.0 &       $-$10.5$\pm$0.6 &      1.13 &      3.08 &              N &        ABD \\
0277-1048331 &   18.57$\pm$0.82~~ &               0.15 &       $-$11.0$\pm$5.8 &       $-$15.6$\pm$3.0 &        $-$8.5$\pm$2.8 &      0.03 &      1.14 &              Y &       BPMG \\
0601-0968580 &   12.83$\pm$1.26~~ &               0.08 &        $-$7.9$\pm$0.5 &       $-$15.9$\pm$2.2 &        $-$9.5$\pm$1.9 &      0.34 &      2.19 &              Y &       BPMG \\
0438-0823866 &   14.32$\pm$0.83~~ &               0.04 &        $-$7.3$\pm$1.2 &       $-$16.4$\pm$1.3 &       $-$10.3$\pm$1.1 &      0.67 &      3.06 &              N &       BPMG \\
0577-1232093 &   19.21$\pm$1.19~~ &              -0.09 &        $-$9.3$\pm$1.2 &       $-$15.0$\pm$1.2 &       $-$10.5$\pm$1.2 &      0.36 &      1.71 &              N &       BPMG \\
\hline                                                                          
\end{tabular}                                                                   
\end{center}                                                                    
\end{table*}

%\input{accretion.tex}
%%%%%%%%%%%%%%%%%%%%%%%%%%%%%%%%%%%%%%%%%%%%%%%%%%%%%%%%%%%%%%%%%%%%%%%%%%%
\subsection{Search for accretion signatures in late-type 
candidates}

A few late-type members of the youngest 
nearby moving groups are still accreting \citep{jayawardhana2006}. 
Therefore, we investigated whether any of our candidates show 
signatures of accretion.
In the spectra of eleven stars, the H$_\alpha$ line was detected in emission.
\citet{barrado2003} proposed an empirical criterion to distinguish 
between stars with chromospheric activity and objects with 
accretion based on their measured H$_\alpha$ emission. 
\citet{whitebasri} used the full width of this line 
at 10\% of the peak to identify "accretors" by 
setting a threshold value of 270 km~s$^{-1}$ for this parameter.   
Using these indicators, we concluded that none of our targets 
show accretion signature and their measured relatively 
weak H$_\alpha$ emission may be of chromospheric origin.

%\input{notes.tex}
%%%%%%%%%%%%%%%%%%%%%%%%%%%%%%%%%%%%%%%%%%%%%%%%%%%%%%%%%%%%%%%%%%%%%%%%%%%

\subsection{Notes on individual targets} \label{notes}

\subsubsection{New members, probable new members}

\paragraph*{\bf 0367-0004557 (HD~3888) and 0361-0008966 (HD~8077):}
Both stars have trigonometric parallaxes in the {\sl Hipparcos} 
catalogue. Their {G}alactic space motion is consistent with 
the characteristic space motion of COL. 
Since their high fractional X-ray luminosity and lithium content 
are also consistent with the respective properties of known members, 
we classified them as new secure members 
of the Columba moving group. 
With its distance of $\sim$45~pc, 0367-0004557 is one of the closest 
members of COL.

\paragraph*{\bf 0540-0027893 (CD$-$36 1289):}
From our analysis, this F8-type star seems to be a probable member of 
COL. {We also} found that 0540-0027893 is a co-moving pair 
of HD~21423 and HD~21434 with a separation of 
85\arcsec and 107\arcsec.
The proper motion of 0540-0027893 
($\mu_{\rm {\alpha}\cos{\delta}}=37.8\pm1.4$\,mas~yr$^{-1}$, 
$\mu_{\rm \delta}=-3.6\pm1.4$\,mas~yr$^{-1}$, NOMAD)
is in good agreement 
with the corresponding astrometric properties of HD~21423 and HD~21434,
$\mu_{\rm {\alpha}\cos{\delta}}=36.3\pm0.5$\,mas~yr$^{-1}$, 
$\mu_{\rm \delta}=-4.7\pm0.5$\,mas~yr$^{-1}$ and 
$\mu_{\rm {\alpha}\cos{\delta}}=34.2\pm1.1$\,mas~yr$^{-1}$, 
$\mu_{\rm \delta}=-5.0\pm1.1$\,mas~yr$^{-1}$, respectively (also from 
 NOMAD). \citet{grenier1999} obtained a radial velocity of 
 18.68~km~s$^{-1}$ for HD~21423 which is identical to the measured RV of 
 18.8$\pm$0.8~km~s$^{-1}$ for our target. The ${\sl Hipparcos}$ parallax 
 of HD~21423 also matches perfectly the kinematic parallax of 
 0540-0027893 (9.27$\pm$0.42~mas~yr$^{-1}$ and 10.06$\pm$0.72~mas~yr$^{-1}$, respectively).
 By recomputing the U,V,W values for 0540-0027893 using the measured 
 trigonometric parallax of HD~21423 instead of the kinematic one, 
 we found that the new values of U = $-$13.7$\pm$0.8, 
 V = $-$22.7$\pm$1.0, W = $-$4.7$\pm$0.9~km~s$^{-1}$ are also 
 very similar to the UVW values of COL (the recomputed $\chi_r^2$ and 
 $d_{UVW}$ values are 0.42 and 1.6~km~s$^{-1}$). 
 Taking {all these} into account, it is likely that 
 HD~21423, HD~21434, and 0540-0027893 constitute a widely 
 separated multiple system and they are secure members of COL. 
HD~21423 has a spectral type of A1V, 
while HD~21434 is an A9V-type star, according to SIMBAD. 
HD~21434 itself is a close binary
with a separation of 0\farcs2 \citep{mason}.
At a parallax of 9.27~mas~yr$^{-1}$, the minimum separation of 
0540-0027893 to HD~21423 and  HD~21434
would be $\sim$9170\,AU and $\sim$11540\,AU, respectively.

\paragraph*{\bf 0454-0034540 (CD$--$44 1533):}
This star was observed in the framework of the SACY survey \citep{torres2006} 
and identified as a K0 dwarf with a $v \sin i$ of  
34$\pm$0.5~km~s$^{-1}$. We classified it as a probable member of COL 
based on its measured properties. 
The high $v \sin i$ value also supports the youth of this target.
Its radial velocity, measured in the $RAVE$ survey \citep{siebert2011}, $v \sin i = $
19.8$\pm$2.0~km~s$^{-1}$ is well consistent with the SACY-based 
RV value of +19.3$\pm$0.2~km~s$^{-1}$. 
 
\paragraph*{\bf 0466-0044706 and 0466-0044732 
(HIP~25434 and HIP~25436):} 
%Based on our analysis these stars were 
We identified these stars as secure members of COL. 0466-0044706 was classified 
as a G0-type star, while 0466-0044732 has a spectral type of F4IV/V.
According to \citet{mason}, 0466-0044732 (component A) and 0466-0044706 (B)
constitute a visual binary system with a separation of 12\arcsec. 
Based on the NOMAD catalogue the proper motion of B 
($\mu_{\rm {\alpha}\cos{\delta}}=23.8\pm1.8$\,mas~yr$^{-1}$, 
$\mu_{\rm \delta}=12.4\pm1.8$\,mas~yr$^{-1}$) is consistent with 
that of A 
($\mu_{\rm {\alpha}\cos{\delta}}=21.4\pm2.3$\,mas, 
$\mu_{\rm \delta}=13.2\pm2.7$\,mas). 
Their {\sl Hipparcos} parallaxes of 14.76$\pm$3.76~mas~yr$^{-1}$ for B and 
13.99$\pm$1.48~mas~yr$^{-1}$ for A agree well within the quoted  
uncertainties. Our RV values of 23.6$\pm$1.8~km~s$^{-1}$  and 
22.6$\pm$6.7~km~s$^{-1}$ for B and A, respectively, 
seems to be consistent as well.  

\citet{desidera2006} also obtained high-resolution spectra for these 
stars and reported RVs of 24.94$\pm$0.33~km~s$^{-1}$  and 
32.17$\pm$1.3~km~s$^{-1}$ for the B and A components, respectively.  
As an explanation {for} the observed RV difference, they suggested that
some of the components might host a further companion.
According to this work, the G0-type B component shows strong chromospheric 
activity ($\log{R^{'}_{HK}=-4.3}$). This large ${R^{'}_{HK}}$ value, together with the measured 
high projected rotational velocities ( 110 and 40~km~s$^{-1}$ for the A and B 
components, respectively) also supports the young age of these stars.

\paragraph*{\bf 0211-0114754 (HD~269620):}
This G6V-type star was also assigned to COL. 
The lithium equivalent width and the radial velocity of the star 
were taken from the SACY catalogue \citep{torres2006} that also 
provided a $v \sin i$ value of 22.3$\pm$3.1~km~s$^{-1}$ for this 
object. 

\paragraph*{\bf 0211-0129478 (HD~269921):}
This G7V-type star is located relatively close to 0211-0114754 in the sky (separation $\sim$49{\arcmin}). It was also 
observed in the framework of {the} SACY survey. This star is a fast 
rotator \citep[$v \sin i$=70~km~s$^{-1}$,][]{torres2006}. 
It was assigned to COL as a probable member in our analysis. 
  
\paragraph*{\bf 0397-0143387 (TYC 8174-1586-1):}
The proper motions ($\mu_{\rm {\alpha}\cos{\delta}}=-185.6\pm1.3$\,mas~yr$^{-1}$, 
$\mu_{\rm \delta}=-99.2\pm2.5$\,mas~yr$^{-1}$ ) 
included in the TYCHO2 catalogue \citep{hog2000}
and proper motions ($\mu_{\rm {\alpha}\cos{\delta}}=-19.8\pm4.5$\,mas~yr$^{-1}$, 
$\mu_{\rm \delta}=14.8\pm4.6$\,mas~yr$^{-1}$ ) from the NOMAD catalogue 
\citep[that use the UCAC2 catalogue data][for this object]{zacharias2004b} 
significantly differ, raising the question which astrometric data is correct.
The NOMAD data are consistent with the proper motion data 
of the UCAC4 \citep[$\mu_{\rm {\alpha}\cos{\delta}}=-15.9\pm8.3$\,mas~yr$^{-1}$, 
$\mu_{\rm \delta}=10.7\pm2.3$\,mas~yr$^{-1}$,][]{zacharias2013} and of the SPM4 
\citep[$\mu_{\rm {\alpha}\cos{\delta}}=-19.45\pm2.6$\,mas~yr$^{-1}$, $\mu_{\rm \delta}=7.42\pm2.61$\,mas~yr$^{-1}$,][]{girard2011} catalogues. 
However, we note that the latter catalogues are somewhat 
dependent on each other e.g. SPM4 and UCAC4 
share some first epoch data \citep{zacharias2013}. 
%thus based only on this consistency  
%we cannot completely exclude TBD.
 According to \citet{torres2006}, 0397-0143387 is a K5eV type star. Assuming that 
it is a normal main-sequence star, its 
absolute V-band magnitude is $M_V \sim$ 7.35 based on \citet{drilling2000}. Using this absolute magnitude 
to derive a photometric distance for the object, and adopting the TYCHO astrometric data 
and the RV taken from \citet{torres2006}, we computed the
 U, V, and W galactic space velocities of 
$-$21.0, $-$21.0, $-$80.7 kms$^{-1}$, respectively. Both the high lithium content (EW$_{\rm Li}$ = 340~m\AA) 
and large X-ray fractional luminosity ($\log{L_{x}/L_{bol}} = -3.18$) of 0397-0143387 
indicate that this star is probably younger than 100 Myr. According to \citet{leggett1992}, 
young disk population stars occupy a well defined region in the 
3-dimensional U, V, W space. The U, V, W velocities obtained from TYCHO 
proper motion data are clearly {outside} this box. {The} obtained large 
velocity dispersion would rather imply a quite old star. Therefore, we consider the
NOMAD/UCAC4/SPM4 data more reliable in this case, {i.e.} this star has a space velocity 
consistent with that of the Carina group.

\paragraph*{\bf 0277-1048331 (HD~173167):}
This F5V-type star was found to be a probable member of
BPMG. It is located close 
to TYC~9073-0762 (with a separation of $\sim$550\arcsec) that was 
proposed by \citet{torres2006} to be a member of BPMG.
These two stars have very similar proper motions, 
$\mu_{\rm {\alpha}\cos{\delta}}=16.1\pm1.3$\,mas~yr$^{-1}$, 
$\mu_{\rm \delta}=-80.2\pm1.3$\,mas~yr$^{-1}$ for 0277-1048331 and 
$\mu_{\rm {\alpha}\cos{\delta}}=13.7\pm3.3$\,mas~yr$^{-1}$, 
$\mu_{\rm \delta}=-82.0\pm3.1$\,mas~yr$^{-1}$ for TYC~9073-0762 (both from NOMAD), 
suggesting 
that they could be companions forming a wide separation 
binary. We found that 0277-1048331 may be a spectroscopic binary 
with a measured systemic radial velocity of +0.8$\pm$7.0~km~s$^{-1}$ 
which is in agreement with the RV of +2.4$\pm$0.1~km~s$^{-1}$ 
measured at TYC~9073-0762 in the SACY survey, however, the uncertainty of 
the former value is quite high.
The estimated kinematic distance of this system is 54~pc 
both based on our analysis and \citet{torres2006}. 
According to \citet{torres2006}, TYC~9073-0762 is an M1Ve type star 
with a very high EW$_{\rm Li}$ of 332~m\AA.

\paragraph*{\bf 0601-0968580:} 
Although 
lithium was not detected in its spectrum, 
this star was classified as a probable member of BPMG 
because it may form a binary system with 
TYC 6872-1011-1, a previously identified member of BPMG \citep{torres2006}. 
%The latter star was proposed by \citet{torres2006} as a member of 
%this group. 
They measured an RV and $v \sin i$ of +4.9$\pm$1.0~km~s$^{-1}$ 
and 33.8$\pm$3.4~km~s$^{-1}$, respectively, and identified 
lithium absorption in the spectrum with an equivalent width of 483~m\AA. 
The separation between 0601-0968580 and TYC 6872-1011-1 is only 
28\farcs3 and both the measured radial velocities ($-$7.0$\pm$0.3~km~s$^{-1}$ for 
0601-0968580) and the proper motion vectors 
($\mu_{\rm {\alpha}\cos{\delta}}=12.9\pm4.4$\,mas~yr$^{-1}$, 
$\mu_{\rm \delta}=-49.4\pm4.4$\,mas~yr$^{-1}$ for 0601-0968580 and 
$\mu_{\rm {\alpha}\cos{\delta}}=6.7\pm2.9$\,mas~yr$^{-1}$, 
$\mu_{\rm \delta}=-42.9\pm2.8$\,mas~yr$^{-1}$ for TYC 6872-1011-1) are quite consistent with
each other, suggesting that the two stars constitute a physical binary system.
 
\paragraph*{\bf 0438-0823866 (2MASS J19233820-4606316):} 
We assigned this M0-type star to BPMG as a probable new member. It has an 
EW$_{\rm Li}$=422$\pm$12~{m\AA}~ 
 (this work) 
confirming that its age is similar to the {kinematic group assigned}. 
\citet{malo2013} also found it to be a probable candidate of BPMG.
Their measured radial velocity of $-$0.7$\pm$0.7~km~s$^{-1}$ is in good {agreement} 
 with our radial velocity value of $-$0.2$\pm$1.1~km~s$^{-1}$.

\paragraph*{\bf 0577-1232093 (2MASS J20055640-3216591):} Our analysis 
identified 0577-1232093 as a probable new member of BPMG. Lithium absorption line was clearly
detected in its spectrum with an EW$_{\rm Li}$=140$\pm$16~{m\AA} that is quite consistent 
with the lithium equivalent widths measured in known BPMG members with similar colour indices 
(Fig.~\ref{ewli}). 
 
\subsubsection{Other targets (rejected candidates)} \label{othnotes}

\paragraph*{\bf 0298-0039269 (HD~37402):} 
Although this F6V-type star may 
{be as old as the Columba group according to its}
%share a common age with COL based on its measured 
lithium content and strong X-ray activity,
we do not classify it as 
a member because it does not entirely fulfill our kinematic 
criteria. Its $\chi_r^2$ is 5.2, lower than the threshold value of 6,
 but its $d_{UVW}=5.3$~km~s$^{-1}$ 
is somewhat larger than the threshold value of $d_{UVW}=$5.0~km~s$^{-1}$.  
{Additional,} more accurate observations of radial velocity and 
astrometric parameters are needed to better assess its membership. 
We note that the star is included in the {\sl Hipparcos} catalogue.   

\paragraph*{\bf 0348-0138354 (TYC 8587-1015-1), 0393-0168117:}
We detected lithium absorption line in the spectra of both targets.
On the one hand the large measured equivalent widths of 252$\pm$14~m\AA~ for 0348-0138354 
and of 277$\pm$10~m\AA~ for 0393-0168117 with their high measured X-ray fractional luminosities 
(Table~\ref{table3}) suggest an age of $<$50~Myr for these targets. 
On the other hand, their {G}alactic space motion does not seem to match the space motion of 
known young nearby moving groups.  However, 0348-0138354 was found to be a spectroscopic 
binary in our analysis implying that its radial velocity measurement is less reliable. 
Further spectroscopic observations are needed to clarify whether the rejection of this candidate was correct or not.  

\paragraph*{\bf 0405-0785226 (2MASS J18151564-4927472):}
Although its kinematic properties 
match those of BPMG, we do not list it as 
a probable member due to the lack of detectable amount of lithium
in the spectrum. We cannot state that it is very young ($\sim$10-20\,Myr) based {solely} on its high 
fractional X-ray luminosity.
%based only on its high fractional X-ray luminosity{\bf,} 
%we cannot establish that it is very young ($\sim$10-20\,Myr).
We note that \citet{malo2013} also identified it as a candidate member of BPMG 
(see their table~5).

\paragraph*{\bf 0706-0872415 (2MASS J21100535-1919573):}
Both the galactic space motion and space position 
of the source are found to be consistent with the respective properties 
of BPMG, however, based on our current age diagnostic, the co-evality 
of the system with BPMG is not satisfactorily established. Therefore 
it has not been assigned as a probable member of BPMG yet. 
This star was also classified as a candidate member of BPMG 
in \citet{malo2013}.

%\input{summary.tex}
%%%%%%%%%%%%%%%%%%%%%%%%%%%%%%%%%%%%%%%%%%%%%%%%%%%%%%%%%%%%%%%%%%%%%%%%%%%%%%%%
\section{Summary} \label{summary}
With the aim to explore the stellar content of five nearby young moving groups 
($\beta$ Pic, Tucana-Horologium, Columba, Carina, Argus) 
further,
we searched for new members in these kinematic assemblages, 
by combining astrometric, photometric, X-ray, and high-resolution 
spectroscopic data. 
Initially, we used the proper motion based 
algorithm proposed by \citet{lepine2009} to identify those stars from a 
cross-correlated NOMAD/{ROSAT} catalogues that have proper motions 
consistent with membership of any of these groups.
This list was 
further refined by selecting only those  
targets whose 
fractional X-ray luminosity and position in the colour-magnitude diagram were 
consistent with those of the known 
members. As a result of this selection algorithm,
we finally identified more than 100 stars that could be potentially 
assigned to at least one of the selected five young moving groups.

In order to further assess their membership status, 
we obtained high-resolution spectra for fifty candidate stars 
using the Echelle spectrograph
at the 2.3-m telescope of the Australian National University 
and the FEROS instrument at the 2.2\,m MPG/ESO telescope.
For four additional objects, we 
used data from the SACY survey \citep{torres2006} 
to complement our observed sample.
Radial velocity data and lithium equivalent widths 
extracted from these spectra offered us the opportunity  
to perform a more accurate kinematic
and age analysis of our candidates and to select 
 targets with kinematic properties and ages 
 consistent with those of the specific groups. 

Based on this analysis, we identified 33 new probable young members of BPMG (4 stars), 
COL (11 stars), CAR (14 stars), and ARG (4 stars) kinematic groups, {as well as 
two new possible members of CAR}.
{In addition} we found serendipitiously a new AB~Dor moving group member.
Trigonometric parallax measurements are needed to securely confirm the 
membership status of these stars. 
For four COL stars (HD~3888, HD~8077, HIP~25434, HIP~25436) 
{\sl Hipparcos} based parallaxes 
have already been available, thus they can be considered 
as secure new members.   
Moreover, we found that 0540-0027893, which is also assigned to COL,  
may be located in 
a wide multiple system. HD~21423, the most massive companion in this 
multiple system, has a {\sl Hipparcos} parallax, which is 
perfectly consistent with the predicted kinematic parallax of
 0540-0027893, making 
the membership of this system 
in Columba to be secure. We note that the membership of 
HD~21423 and HD~21434, the two early-type (A1V and A9V) 
companions is also a new discovery. 
Two new probable BPMG members, 
0277-1048331 and 
0601-0968580 seem to form binary systems with two 
previously identified members {of this group} (TYC~9073-0762 and 
TYC 6872-1011-1, respectively). 
We significantly increased the number of known probable/possible members in the Carina group 
by adding 16 new stars to {those} 23 already listed 
in \citet{torres2008}.

Our preselection method turned out to be very successful. From the 54 
candidates, 34 were verified as probable young moving group members 
based on the analysis of high-resolution spectroscopic data.

\section*{Acknowledgements}
We thank our anonymous referee whose comments significantly improved the manuscript.
This project was supported by the Hungarian OTKA grants K76816, K83790, K101393, K104607, 
the HUMAN MB08C 81013 project of the MAG Zrt., the PECS-98073 program of the European Space 
Agency (ESA) and the Hungarian Space Office, the Lend\"ulet 2009/2012 Young Researchers'
Programs and the Bolyai Research Fellowship of the Hungarian Academy of Sciences.
This publication makes use of data products from the Wide-field Infrared Survey Explorer, 
which is a joint project of the University of California, Los Angeles, and the 
Jet Propulsion Laboratory/California Institute of Technology, funded by the National 
Aeronautics and Space Administration.
The publication also makes use of data products from the Two Micron All Sky Survey, which
is a joint project of the University of Massachusetts and the Infrared Processing and
Analysis Center/California Institute of Technology, funded by the National Aeronautics
and Space Administration and the National Science Foundation.
Our research has made use of the VizieR catalogue access tool, 
CDS, Strasbourg, France.

\appendix 
\section{Description of the selected five young associations} \label{appendix}

\paragraph*{BPMG} was first proposed by \citet{zuckerman2001}. 
After \citet{barrado1999} found two young M-type proper motion companions to $\beta$~Pic, 
\citet{zuckerman2001} defined this group by identifying 18 systems that likely belong to it. 
In their review work, \citet{torres2008} listed 48 high probability members, most of them 
concentrated in the southern hemisphere, {with distances ranging} between 9 and 80~pc.
Thanks to recent studies \citep{lepine2009,rice2010,schlieder2010,kiss2011,moor2011b,schlieder2012b,malo2013},
 the number of probable members has been increased considerably 
in the last few years, with many newly identified 
systems located in the northern
hemisphere \citep[e.g.][]{schlieder2010,schlieder2012b,moor2011b}.  
Based on the members' location in the H-R diagram and from their lithium equivalent widths,
\citet{zuckerman2001} derived an age of 12$^{+8}_{-4}$\,Myr for the group.
Using dynamical back-tracing models of members, \citet{ortega2002} and \citet{song2003} 
obtained very similar age estimates of ~12~Myr for this group.
Recently \citet{mentuch2008} derived a somewhat 
higher age, 21$\pm$9\,Myr, from lithium depletion in the group.
  
\paragraph*{Tucana-Horologium, Columba and Carina moving groups} together constitute the so-called 
Great Austral Young Association \citep[GAYA,][]{torres2008}. 
{Considering} Li abundances, these three groups are indistinguishable, 
indicating almost identical ages \citep{daSilva2009}.
Members of these groups can be separated based on their galactic space motion and galactic 
space location, although the characteristic U,V,W space velocities of COL
are quite similar to those of CAR and the space regions occupied by their members 
also {overlap partially} making the accurate group assignment difficult in some cases.  
THA is the nearest with a mean distance of $\sim$48~pc, and {therefore 
it was discovered first} \citep{zuckwebb2000,torres2000} among the three groups.  
COL and CAR groups were proposed later as separate entities by \citet{torres2008}. 
Prior to that, some COL and CAR members were assigned to the GAYA2 group \citep{torres2003}.  
\citet{torres2008} quoted 44, 53, and 23 probable members of THA, COL, and CAR, respetively.
\citet{daSilva2009} added six new stars to COL, but also rejected three from those listed in \citet{torres2008}.
Recently \citet{zuckerman2011} {and} \citet{zuckerman2012} contributed significantly to the census of THA and COL by 
proposing in total 12 and 17 new members, respectively. 
The majority of these stars are early-type with spectral types A or F.
\citet{kiss2011} and \citet{moor2011b} also identified some additional probable THA and COL members. 
Similarly to BPMG, most members of the GAYA complex are located in South.
The age estimates for THA range between 10 and 40\,Myr 
\citep{zuckerman2001b,makarov2007,mentuch2008,daSilva2009}.
The most widely accepted value at the moment is 30~Myr, which is the proposed age 
of the whole GAYA 
complex \citep{torres2008}.

\paragraph*{Argus association.}
The existence of the Argus association was proposed by \citet{torres2003}. 
ARG comprises both a sparsely distributed "classic" moving group and 
members of the IC~2391 open cluster \citep{torres2008}. 
The 29 members identified in the sparse part of the group are usually found to be closer to us than 
stars that belong to IC~2391 that are all located at a distance $>$120~pc.  
\citet{zuckerman2011} and \citet{zuckerman2012} found 13 new nearby ($<$80~pc) 
early-type Argus members. According to \citet{riedel2011}, the nearest 
pre-main sequence star, AP~Col, with a distance of 8.4~pc can also be assigned to this group. 
 Based on the members' position in the H-R diagram and on their lithium equivalent widths,
\citet{torres2008} derived an age of 40~Myr for ARG. \citet{barrado2004} obtained an age estimate 
 of 50$\pm$5~Myr for IC~2391 based on the position of the ``Li depletion boundary''. 
 Comparing the position of recently proposed A-type members of Argus with the location of 
 Pleiades stars, \citet{zuckerman2012} suggested that the group may be somewhat older than the {previously} 
 determined 
 40~Myr. The Argus association is also concentrated in the Southern hemisphere.

%\section[]{Large gaps in L\lowercase{y}${\balpha}$ forests\\* due 
%to fluctuations in line distribution}

%\bsp

\label{lastpage}


\begin{thebibliography}{99}
\bibitem[\protect\citeauthoryear{Barrado y Navascu{\'e}s et 
al.}{1999}]{barrado1999} Barrado y Navascu{\'e}s D., Stauffer 
J.~R., Song I., Caillault J.-P., 1999, ApJ, 520, L123 
\bibitem[\protect\citeauthoryear{Barrado y Navascu{\'e}s 
\& Mart{\'{\i}}n}{2003}]{barrado2003} Barrado y Navascu{\'e}s D., Mart{\'{\i}}n E.~L., 2003, AJ, 126, 2997
\bibitem[\protect\citeauthoryear{Barrado y Navascu{\'e}s, Stauffer, 
\& Jayawardhana}{2004}]{barrado2004} Barrado y Navascu{\'e}s D., Stauffer J.~R., Jayawardhana R., 2004, ApJ, 614, 386
\bibitem[\protect\citeauthoryear{Carson et al.}{2013}]{carson2012} 
Carson J., et al., 2013, ApJ, 763, L32 
\bibitem[\protect\citeauthoryear{Chambers}{2001}]{chambers2001} 
Chambers J.~E., 2001, Icar, 152, 205 
\bibitem[\protect\citeauthoryear{da Silva et al.}{2009}]{daSilva2009} 
da Silva L., et al., 2009, A\&A, 508, 833 
\bibitem[\protect\citeauthoryear{Desidera et al.}{2006}]{desidera2006} Desidera S., 
Gratton R.~G., Lucatello S., Claudi R.~U., Dall T.~H., 2006, A\&A, 454, 553
\bibitem[\protect\citeauthoryear{Drilling \& Landolt}{2000}]{drilling2000} Drilling J.~S. \& Landolt A.~U., Infrared Astronomy, 
in Cox A. N., ed., Allen's Astrophysical Quantities, 4th edn. AIP Press, New York, ISBN: 0387987460. 2000. p. 388.
\bibitem[\protect\citeauthoryear{Fern{\'a}ndez, Figueras \& Torra}{2008}]{fernandez2008} Fern{\'a}ndez D., Figueras F., Torra J., 2008, A\&A, 480, 735 
\bibitem[\protect\citeauthoryear{Girard et al.}{2011}]{girard2011} 
Girard T.~M., et al., 2011, AJ, 142, 15 
\bibitem[\protect\citeauthoryear{Grenier et al.}{1999}]{grenier1999} Grenier S., Burnage R., Faraggiana R., Gerbaldi M., Delmas F., 
G{\'o}mez A.~E., Sabas V., Sharif L., 1999, A\&AS, 135, 503
\bibitem[\protect\citeauthoryear{Hauschildt, Allard, 
\& Baron}{1999}]{hauschildt} Hauschildt P.~H., Allard F., Baron E., 1999, ApJ, 512, 377
\bibitem[\protect\citeauthoryear{H{\o}g et al.}{2000}]{hog2000} H{\o}g E., et al., 2000, A\&A, 355, L27 
\bibitem[\protect\citeauthoryear{Jayawardhana et 
al.}{2006}]{jayawardhana2006} Jayawardhana R., Coffey J., Scholz A., 
Brandeker A., van Kerkwijk M.~H., 2006, ApJ, 648, 1206 
\bibitem[\protect\citeauthoryear{Kasper et al.}{2007}]{kasper2007} Kasper M., et al., 2007, A\&A, 472, 321 
\bibitem[\protect\citeauthoryear{Kaufer et al.}{1999}]{kaufer1999} 
Kaufer A., Stahl O., Tubbesing S., N{\o}rregaard P., Avila G., Francois P., 
Pasquini L., Pizzella A., 1999, Msngr, 95, 8 
\bibitem[\protect\citeauthoryear{Kenyon 
\& Bromley}{2006}]{kb2006} Kenyon S.~J., Bromley B.~C., 2006, AJ, 131, 1837
\bibitem[\protect\citeauthoryear{Kenyon 
\& Bromley}{2008}]{kb2008} Kenyon S.~J., Bromley B.~C., 2008, ApJS, 179, 451
 \bibitem[\protect\citeauthoryear{Kiss et al.}{2011}]{kiss2011} 
Kiss L.~L., et al., 2011, MNRAS, 411, 117 
\bibitem[\protect\citeauthoryear{Lagrange et 
al.}{2010}]{lagrange2010} Lagrange A.-M., et al., 2010, Sci, 329, 57 
\bibitem[\protect\citeauthoryear{Lallement et 
al.}{2003}]{lallement2003} Lallement R., Welsh B.~Y., Vergely J.~L., Crifo F., Sfeir D., 2003, A\&A, 411, 447
\bibitem[\protect\citeauthoryear{Leggett}{1992}]{leggett1992} Leggett S.~K., 1992, ApJS, 82, 351 
\bibitem[\protect\citeauthoryear{L{\'e}pine 
\& Simon}{2009}]{lepine2009} L{\'e}pine S., Simon M., 2009, AJ, 137, 3632
\bibitem[\protect\citeauthoryear{Makarov}{2007}]{makarov2007} Makarov V.~V., 2007, ApJS, 169, 105
\bibitem[\protect\citeauthoryear{Malo et al.}{2013}]{malo2013} 
Malo L., Doyon R., Lafreni{\`e}re D., Artigau {\'E}., Gagn{\'e} J., Baron 
F., Riedel A., 2013, ApJ, 762, 88 
\bibitem[\protect\citeauthoryear{Marois et al.}{2008}]{marois2008} 
Marois C., Macintosh B., Barman T., Zuckerman B., Song I., Patience J., 
Lafreni{\`e}re D., Doyon R., 2008, Sci, 322, 1348 
%\bibitem[\protect\citeauthoryear{Mamajek 
%\& Hillenbrand}{2008}]{mamajek2008} Mamajek E.~E., Hillenbrand L.~A., 2008, ApJ, 687, 1264 
\bibitem[\protect\citeauthoryear{Mason et al.}{2001}]{mason} 
Mason B.~D., Wycoff G.~L., Hartkopf W.~I., Douglass G.~G., Worley C.~E., 
2001, AJ, 122, 3466 
\bibitem[\protect\citeauthoryear{Mentuch et al.}{2008}]{mentuch2008} Mentuch E., Brandeker A., van Kerkwijk 
M.~H., Jayawardhana R., Hauschildt P.~H., 2008, ApJ, 689, 1127 
%\bibitem[\protect\citeauthoryear{Mo{\'o}r et al.}{2011}]{moor2011} 
%Mo{\'o}r A., et al., 2011, ApJ, 740, L7 
\bibitem[\protect\citeauthoryear{Mo{\'o}r et al.}{2011}]{moor2011b} 
Mo{\'o}r A., et al., 2011, ApJS, 193, 4
\bibitem[\protect\citeauthoryear{Munari et 
al.}{2005}]{munari2005} Munari U., Sordo R., Castelli F., Zwitter T., 2005, A\&A, 442, 1127
\bibitem[\protect\citeauthoryear{Ortega et al.}{2002}]{ortega2002} 
Ortega V.~G., de la Reza R., Jilinski E., Bazzanella B., 2002, ApJ, 575, L75 
\bibitem[\protect\citeauthoryear{Pascucci et 
al.}{2006}]{pascucci2006} Pascucci I., et al., 2006, ApJ, 651, 1177
\bibitem[\protect\citeauthoryear{Pecaut \& Mamajek}{2013}]{pecaut2013} Pecaut M.~J., Mamajek E.~E., 2013, ApJ, accepted, arXiv:1307.2657 
\bibitem[\protect\citeauthoryear{Rice, Faherty, 
\& Cruz}{2010}]{rice2010} Rice E.~L., Faherty J.~K., Cruz K.~L., 2010, ApJ, 715, L165 
\bibitem[\protect\citeauthoryear{Riedel et al.}{2011}]{riedel2011} 
Riedel A.~R., Murphy S.~J., Henry T.~J., Melis C., Jao W.-C., Subasavage 
J.~P., 2011, AJ, 142, 104 
\bibitem[\protect\citeauthoryear{Schlieder, L{\'e}pine, 
\& Simon}{2010}]{schlieder2010} Schlieder J.~E., L{\'e}pine S., Simon M., 2010, AJ, 140, 119
\bibitem[\protect\citeauthoryear{Schlieder, L{\'e}pine, 
\& Simon}{2012a}]{schlieder2012a} Schlieder J.~E., L{\'e}pine S., Simon M., 2012, AJ, 143, 80 
 \bibitem[\protect\citeauthoryear{Schlieder, L{\'e}pine, 
\& Simon}{2012b}]{schlieder2012b} Schlieder J.~E., L{\'e}pine S., Simon M., 2012, AJ, 144, 109  
\bibitem[\protect\citeauthoryear{Shkolnik, Liu, \& Reid}{2009}]{shkolnik2009} 
Shkolnik E., Liu M.~C., Reid I.~N., 2009, ApJ, 699, 649 
\bibitem[\protect\citeauthoryear{Shkolnik et 
al.}{2011}]{shkolnik2011} Shkolnik E.~L., Liu M.~C., Reid I.~N., 
Dupuy T., Weinberger A.~J., 2011, ApJ, 727, 6 
\bibitem[\protect\citeauthoryear{Shkolnik et al.}{2012}]{shkolnik2012} Shkolnik E.~L., Anglada-Escud{\'e} G., Liu 
M.~C., Bowler B.~P., Weinberger A.~J., Boss A.~P., Reid I.~N., Tamura M., 
2012, ApJ, 758, 56
\bibitem[\protect\citeauthoryear{Siebert et 
al.}{2011}]{siebert2011} Siebert A., et al., 2011, AJ, 141, 187  
\bibitem[\protect\citeauthoryear{Skrutskie et 
al.}{2006}]{skrutskie} Skrutskie M.~F., et al., 2006, AJ, 131, 
1163 
\bibitem[\protect\citeauthoryear{Song, Zuckerman, 
\& Bessell}{2003}]{song2003} Song I., Zuckerman B., Bessell M.~S., 2003, ApJ, 599, 342 
\bibitem[\protect\citeauthoryear{Torres et al.}{2000}]{torres2000} 
Torres C.~A.~O., da Silva L., Quast G.~R., de la Reza R., Jilinski E., 
2000, AJ, 120, 1410
\bibitem[\protect\citeauthoryear{Torres et al.}{2003}]{torres2003} 
Torres C.~A.~O., Quast G.~R., de La Reza R., da Silva L., Melo C.~H.~F., 
Sterzik M., 2003, ASSL, 299, 83  
\bibitem[\protect\citeauthoryear{Torres et al.}{2006}]{torres2006} 
Torres C.~A.~O., et al., 2006, A\&A, 460, 695 
\bibitem[\protect\citeauthoryear{Torres et al.}{2008}]{torres2008} 
Torres C.~A.~O., et al., 2008, Handbook of Star Forming Regions, Vol. II, 
ASP Monograph Publications, Vol. 5, 757
\bibitem[\protect\citeauthoryear{Zacharias et al.}{2004a}]{zacharias2004} Zacharias N., Monet D.~G., Levine S.~E., 
Urban S.~E., Gaume R., Wycoff G.~L., 2004, AAS, 36, 1418 
\bibitem[\protect\citeauthoryear{Zacharias et al.}{2004b}]{zacharias2004b} Zacharias N., Urban S.~E., Zacharias 
M.~I., Wycoff G.~L., Hall D.~M., Monet D.~G., Rafferty T.~J., 2004, AJ, 
127, 3043 
\bibitem[\protect\citeauthoryear{Zacharias et al.}{2013}]{zacharias2013} Zacharias N., Finch C.~T., Girard T.~M., 
Henden A., Bartlett J.~L., Monet D.~G., Zacharias M.~I., 2013, AJ, 145, 44 
\bibitem[\protect\citeauthoryear{Zuckerman 
\& Webb}{2000}]{zuckwebb2000} Zuckerman B., Webb R.~A., 2000, ApJ, 535, 959
\bibitem[\protect\citeauthoryear{Zuckerman et 
al.}{2001}]{zuckerman2001} Zuckerman B., Song I., Bessell M.~S., Webb 
R.~A., 2001, ApJ, 562, L87 
\bibitem[\protect\citeauthoryear{Zuckerman, Song, 
\& Webb}{2001}]{zuckerman2001b} Zuckerman B., Song I., Webb R.~A., 2001, ApJ, 559, 388
\bibitem[\protect\citeauthoryear{Zuckerman 
\& Song}{2004}]{zucksong2004} Zuckerman B., Song I., 2004, ARA\&A, 42, 685
\bibitem[\protect\citeauthoryear{Zuckerman et 
al.}{2011}]{zuckerman2011} Zuckerman B., Rhee J.~H., Song I., Bessell 
M.~S., 2011, ApJ, 732, 61
\bibitem[\protect\citeauthoryear{Zuckerman 
\& Song}{2012}]{zuckerman2012} Zuckerman B., Song I., 2012, ApJ, 758, 77 
\bibitem[\protect\citeauthoryear{van Leeuwen}{2007}]{vanleeuwen07} 
van Leeuwen, F.\ 2007, Astrophysics and Space Science Library, 350, 
by F. van Leeuwen
\bibitem[\protect\citeauthoryear{Voges et 
al.}{1999}]{voges1999} Voges W., et al., 1999, A\&A, 349, 389 
\bibitem[\protect\citeauthoryear{Voges et al.}{2000}]{voges2000} 
Voges W., et al., 2000, IAUC, 7432, 1
\bibitem[\protect\citeauthoryear{White 
\& Basri}{2003}]{whitebasri} White R.~J., Basri G., 2003, ApJ, 582, 1109
\bibitem[\protect\citeauthoryear{Wright et al.}{2010}]{wright} 
Wright E.~L., et al., 2010, AJ, 140, 1868 
\bibitem[\protect\citeauthoryear{Wyatt}{2008}]{wyatt2008} Wyatt M.~C., 2008, ARA\&A, 46, 339 
\end{thebibliography}
\end{document}